\documentclass{emulateapj}
\usepackage{graphicx}
\usepackage{hyperref}

\def \th {\thinspace}

\def \arcmin {\hbox{$^\prime$}}
\def \arcsec {\hbox{$^{\prime\prime}$}}

\def\spose#1{\hbox to 0pt{#1\hss}}
\def\ltsim{$\mathrel{\spose{\lower 3pt\hbox{$\sim$}}
        \raise 2.0pt\hbox{$<$}}$\thinspace}
\def\gtsim{$\mathrel{\spose{\lower 3pt\hbox{$\sim$}}
        \raise 2.0pt\hbox{$>$}}$\thinspace}

\def \nh {${\rm N_{H}}$}
\def \src {NGC\th 1332}
\def \srctwo {NGC\th 1331}
\def \deg {$^\circ$}
\def \eg {e.g.}
\def \ie {i.e.}
\def \cf {c.f.}
\def \dtwentyfive {${\rm D_{25}}$}
\def \chandra {{\em Chandra}}
\def \acis {{\em ACIS}}
\def \ciao {{\em CIAO}}
\def \caldb {{\em Caldb}}
\def \xspec {{\em XSPEC}}
\def \isis {{\em ISIS}}
\def \heasoft {{\em Heasoft}}
\def \ergps {${\rm erg\ s^{-1}}$}

\def \asca {{\em ASCA}}
\def \rosat {{\em Rosat}}
\def \pspc {{\em PSPC}}

\def \lx {${\rm L_X}$}
\def \leda {{\em{LEDA}}}
\def \ned {{\em{NED}}}

\shorttitle{X-ray point sources in NGC\th 1332}
\shortauthors{Humphrey and Buote}

\begin{document}

\title{A Chandra view of the normal S0 galaxy NGC\th 1332: I. An unbroken, steep power law luminosity function for the LMXB population.}
\author {Philip J. Humphrey and  David A. Buote}
\affil{Department of Physics and Astronomy, University of California at Irvine, 4129 Frederick Reines Hall, Irvine, CA 92697-4575}
\begin{abstract}
Chandra \acis-S3 observations of the nearby S0 galaxy
\src\ resolve much of the X-ray emission into 73 point-sources, of which
37 lie within the \dtwentyfive\ isophote. The remaining galaxy emission
comprises hot, diffuse gas and unresolved sources and is
discussed in two companion papers.
The point-source luminosity function (XLF) shows the characteristic break seen
in other early-type galaxies at $\sim 2\times 10^{38}$ \ergps. 
After applying corrections for detection incompleteness at low luminosities
due to source 
confusion and contamination from diffuse galactic emission,
the break vanishes and the data are well-described as a single power law. 
This result casts further doubt on there being a ``universal''
XLF break in early-type galaxies marking the division between
neutron-star and black-hole systems.
The logarithmic slope of the differential XLF (dN/dL), $\beta=2.7\pm0.5$,
is marginally ($\sim2.5\sigma$) 
steeper than has been found for analogous 
completeness-corrected fits of  other early-type galaxies but closely
matches the behaviour seen at
high luminosities in these systems. 
Two of the sources within \dtwentyfive\ are Ultra-luminous X-ray sources (ULX),
although neither have  \lx$>2\times 10^{39}$ \ergps. The absence of 
very luminous ULX in  early-type galaxies suggests a break 
in the XLF slope at $\sim$1--2$\times 10^{39}$ \ergps, although the data were
not of sufficient quality to constrain such a feature in \src. 
The sources have a spatial  distribution consistent with the optical
light and  display a range of characteristics that are consistent with an LMXB 
population. The general spectral characteristics of the individual sources,
as well as the  composite source spectra, are in good agreement with 
observations
of other early-type galaxies, although a small number of highly-absorbed
sources are seen. Two sources have very soft spectra,
two show strong variability,
indicating compact binary nature and one source shows evidence of an 
extended radial profile. We do not detect a central source in \src, but we 
find a faint (\lx$=2\pm1 \times 10^{38}$ \ergps) point-source
coincident with the centre of the companion dwarf galaxy \srctwo. 
\end{abstract}

\keywords{galaxies: elliptical and lenticular, cD --- galaxies: individual (NGC 1332) --- X-rays: binaries --- X-rays: galaxies}

\section{Introduction}
Prior to the launch of \chandra\ only a small number of the very
brightest point-sources in early type galaxies could be 
resolved from the diffuse galactic emission
\citep[][]{fabbiano89,colbert99,roberts00}. The study of their point-source 
populations was therefore restricted to the average, 
composite properties inferred by decomposing the
emission into a number of spectral or spatial model components 
\citep[\eg][]{matsushita94,brown01}. The advent of \chandra\ has 
revolutionized this field, however, allowing a large fraction of the 
sources to be resolved and studied directly 
\citep[for a recent review, see][]{fabbiano03}. Extragalactic point-sources 
most probably represent an heterogeneous mixture of 
different source types, although the old stellar populations characteristic of 
early-type galaxies suggest a predominantly low-mass X-ray binary (LMXB) 
nature. As endpoints of stellar 
evolution, the properties of X-ray binaries are a crucial
diagnostic for the evolution of the stars as a whole within the galaxy.
Although the data are generally of insufficient quality to investigate
fully for each source the rich, diagnostic phenomenology of Galactic LMXB 
\citep{white95}, in some cases it has been possible to identify 
variability \citep[\eg][]{sarazin01,kraft01} and the 
spectral signatures of black-hole binaries 
\citep[\eg][]{makishima00,humphrey03a}.

In order to study the properties of the X-ray point-source population
as a whole, it is common practice to consider the X-ray 
luminosity function (XLF)
\citep[\eg][]{sarazin01,blanton01b,zezas02d}, the shape of which 
is a strong function of the age of the stellar population 
\citep{kilgard02,belczynski03}. 
Between early-type galaxies, there is remarkable similarity in the shape of 
the XLF \citep{kim03a}, which is typically found to be
a steep power law with a break occurring around 2--4 $\times 10^{38}$ \ergps\
and a high-luminosity slope, $\beta \simeq$ 
2--3 \citep{sarazin01,blanton01b,kraft01,colbert03a}. 
These slopes are also broadly consistent with 
old stellar populations in M31 \citep{kong03}.
It has been suggested that the presence of breaks at luminosities at 
around the Eddington limit of a 1.4${\rm M_\odot}$ neutron star 
(${\rm L_{EDD}=}$2--4$\times 10^{38}$ \ergps, depending on the composition of 
the accreting matter and the neutron-star equation of state:
\citealt{paczynski83}) may arise from  a division between neutron-star 
and black-hole systems \citep[\eg][]{sarazin01}.
However in the elliptical galaxy NGC\th 720, a substantially higher 
luminosity break, at ${\rm \sim 1\times 10^{39}}$ \ergps\ was 
found instead by \citet{jeltema03}.
\citet{kim03b} pointed out that in the elliptical galaxy NGC\th 1316,
a break is arises only if no correction is made for  point-source 
detection incompleteness. Applying this correction to a sample of 
early-type galaxies, \citet{kim03a} found a similar result,
although they still found marginal evidence of a break
at $\sim 5 \times 10^{38}$ \ergps.
This luminosity is
sufficiently high, however, to cast further doubt upon
the idea that it marks the division between neutron-star and black-hole
binaries.

At the extreme end of the XLF are the  so-called 
``Ultra Luminous X-ray sources'' (ULX), which are 
non-nuclear objects with luminosities exceeding $10^{39}$ \ergps, sometimes
reaching as high as $10^{41}$ \ergps\
\citep[\eg][]{fabbiano89,makishima00,zezas02b,davis03}. They have been
found in galaxies of all morphological types \citep{colbert02},
although there is a strong association between ULX and star-formation
so that their specific frequency is far lower in early-type galaxies
\citep{kilgard02,humphrey03a,irwin03b}. However, in the otherwise normal
elliptical NGC\th 720, \citet{jeltema03} found a remarkable population of 
9 ULXes. The nature
of ULX is somewhat enigmatic. Although the less extreme objects
(\lx$\sim$1--2$\times 10^{39}$ \ergps) are consistent with isotropic emission 
from Eddington-limited black-hole binaries with compact-object masses
$\sim$10--20${\rm M_\odot}$, there has been much debate over whether the
most luminous objects represent ``Intermediate mass black holes''
with mases \gtsim ${\rm 100 M_\odot}$, super-Eddington emission from
sources in the thermal-timescale mass-transfer phase, or beamed emission 
from lower-luminosity sources \citep{colbert99,king01,king02}.

Another feature of the LMXB population of particular interest is its
possible association with globular clusters (GCs), since it provides
valuable insight into X-ray binary formation. The fraction of the 
extragalactic LMXB identified with GCs varies considerably between
galaxies, from
$\sim$20--70\% \citep{angelini01,kundu02a,sarazin01}. Nonetheless,
there is evidence that the frequency of at least part of the LMXB 
population correlates with the numbers of GCs so that 
$\sim$4\% of all GCs contain an LMXB, consistent with the Milky Way 
\citep{kundu03,sarazin03}. These
authors also found evidence that LMXB favour brighter, redder GCs. 

In this paper, we analyse with \chandra\ the hitherto unresolved point-source
population of the lenticular galaxy \src. Using \rosat\ \pspc\ data of 
this galaxy, 
\citet[][hereafter BC]{buote96a} found elongation of the X-ray 
isophotes and demonstrated that the galaxy mass profile was marginally 
inconsistent with the optical light, indicating the presence of a substantial 
dark matter halo. Unfortunately, most of the point-source population
could not be resolved, so that it was necessary to estimate limits for the 
unresolved point-source contribution to the X-ray emission, and 
tight constraints on the shape of the dark matter halo could not be obtained.
Using \asca\ data \citet{buote97a} found a hard spectral component in the 
integrated stellar emission, indicating a significant contribution from
unresolved LMXB, although this did not allow these sources to be studied in
detail.\chandra\ allows us for the first time to separate a large fraction
of the LMXB from the diffuse emission of \src, and to study them 
directly. The properties of the  diffuse gas and the mass distribution
are discussed separately in two companion papers 
\citep[][hereafter Papers I and II, respectively]{humphrey04b,buote04a}.

In order to determine the properties of the stellar population 
accurately, it is important to adopt a reliable distance estimate
(\cf\ NGC\th 4038/9, for which a recently revised distance estimate 
would reduce the reported ULX population threefold; \citealt{saviane03}).
For NGC\th 1332, distances determined both by surface brightness fluctuations
(SBF) and the globular cluster luminosity function (GCLF) are in excellent
agreement, being $23\pm3$\th Mpc and $22\pm6$\th Mpc, respectively 
\citep{tonry01,kundu01}.  All errors quoted here, and subsequently, are 90\% confidence limits unless otherwise stated.

\section{Observations and data analysis}

\begin{figure*}
\plottwo{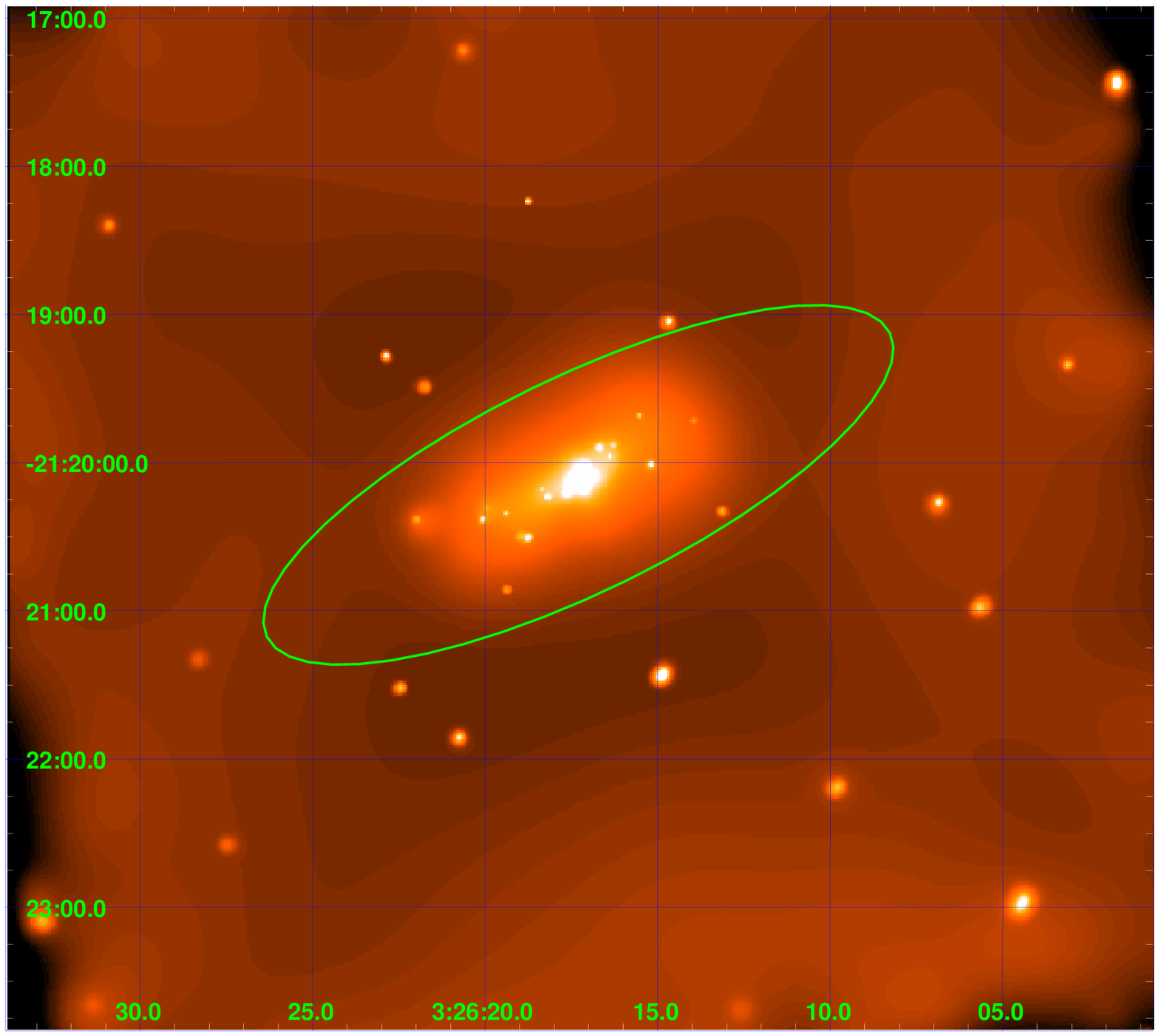}{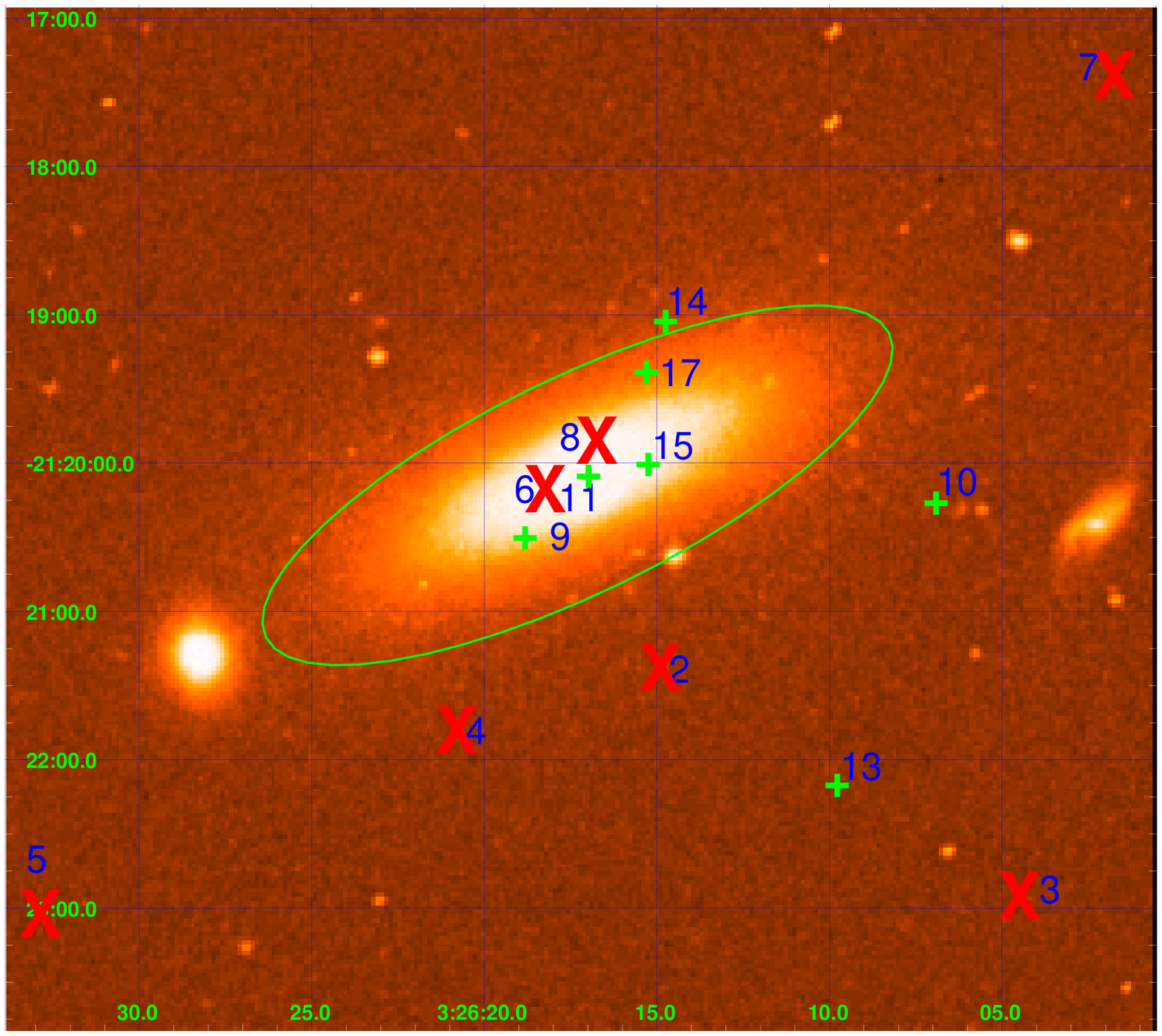}
\caption{left panel: False colour, smoothed image of the central part of the 
S3 chip. The image was initially flat-fielded with an 
exposure-map 
created at 1~keV  (peak energy).Overlaid is the \dtwentyfive\ ellipse taken 
from \citet{devaucouleurs91}, but the position angle has been altered to 
116\deg\ to agree with BC.  Right panel: the DSS optical image,
overlaid with the \dtwentyfive\ ellipse and, following \citet{humphrey03a},
the positions of X-ray sources with \lx$>10^{39}$ \ergps\ (X) and those with 
$5\times 10^{38}<$\lx$<10^{39}$ \ergps\ (+). The identification numbers of a selection of interesting sources are given on the figure. The bright galaxy to the left of
\src\ is NGC\th 1331.} \label{fig_image}
\end{figure*}

The region of sky containing \src\ was observed with the ACIS instrument 
aboard \chandra\ between  2002 September 19 10:39 and September 20 02:59 UTC,
the galaxy being centred on the S3 chip, for a nominal $\sim$60~ks exposure.
An additional observation made between September 18 02:56 and 08:27 UTC,
for a nominal 20~ks exposure,
was heavily contaminated by flaring of the background and so we present
results only for the better-quality dataset.
For data-reduction, we used the \ciao\th~3.0.1 and \heasoft\th~5.2 
suites  of software tools, and for spectral analysis we used
\xspec\ 11.2.0 and \isis\ 1.1.5.
In order to use the most up-to-date calibration, we
reprocessed the Level-1 data, following the standard
\chandra\ threads\footnote{\href{http://cxc.harvard.edu/ciao/threads/index.html}{http://cxc.harvard.edu/ciao/threads/index.html}}. 
We corrected a time-dependent drift in the gain with the  {\tt apply\_gain}\footnote{\href{http://cxc.harvard.edu/cont-soft/software/corr_tgain.1.0.html}{http://cxc.harvard.edu/cont-soft/software/corr\_tgain.1.0.html}}
task. To identify periods of unusually high background
 for which the degradation in signal-to-noise can affect analysis, 
a lightcurve was generated from source-free regions of the active chips and 
examined by eye for ``flares''. For much of the observation there was evidence
of low-level flaring (${\rm \Delta R/R\sim}$20\%, where R is the count-rate). 
For point-source identification and characterisation, we found the
S/N improvement from reducing the background level (by $\sim$10\%)
was outweighed by the  reduction in the exposure-time (by $\sim$30\%)
if we excised data during the strongest flares.
We therefore used the entire observation, giving a total exposure time of 
57~ks. We have systematically confirmed that including the flaring did not bias our results. A more thorough discussion of the flaring is given in Paper II.
Fig.~\ref{fig_image} shows the image of the inner part of the S3 chip,
which has been  flat-fielded with an exposure-map generated with \ciao\
at 1.0~keV 
(approximately the peak energy of the diffuse emission), and  smoothed 
with the \ciao\ task {\tt csmooth} (taking the default
parameter values). A number of bright point-sources are clearly 
visible in the X-ray data. For comparison purposes the  Digitized Sky Survey 
(DSS) optical image is also shown.

\section{Point source detection} \label{sect_detection}
Point-source detection was performed using the \ciao\ tool {\tt wavdetect}
\citep{freeman02}. In order to improve the likelihood of identifying sources
with peculiarly hard or soft spectra, full-resolution 
images were created of the region of the 
\acis\ focal-plane containing the S3 chip in three different energy-bands
(0.1--10.0~keV, 0.1--3.0~keV and 3.0--10.0~keV). Sources were
detected separately in each image. In order to minimize spurious detections at 
node or chip boundaries we supplied the detection algorithm with 
exposure-maps  generated at 
energies 1.7~keV, 1.0~keV and 7~keV respectively (although the precise
energies chosen made little difference to the results). The
detection algorithm searched for structure over pixel-scales of 1, 2, 4, 8 and
16 pixels, and the detection threshold was set to $\sim 10^{-7}$ spurious 
sources per pixel (corresponding to $\sim$0.1 spurious detections per 
image). The source-lists obtained within each energy-band were combined and
duplicated sources removed, and the final list was checked  
by visual inspection of the images. 

A total of 76 sources were detected on the S3 chip, of which the brightest
and most extended was centred  within 1\arcsec\ of the galaxy centroid 
as given in \ned\ (which is accurate to $\sim$ 2\arcsec). Examination of the
azimuthally-averaged radial brightness profile of this ``source'' 
(Sect.~\ref{sect_srcextent}) clearly indicated it to be very extended and
so we identified this with the central part of the  diffuse galactic emission 
(which had been spuriously identified by {\tt wavdetect} 
as a point-source due to 
its being very centrally-peaked). We found that one point-source lay within
the B-band 25${\rm ^{th}}$ magnitude (\dtwentyfive) isophote
of NGC\th 1331 (as listed in \citealt{devaucouleurs91}). In fact, this source
was within 1\arcsec\ of the galaxy centroid listed in \ned, but was
rather faint (with a count-rate ${\rm \sim 4\times 10^{-4} count\ s^{-1}}$, 
corresponding to ${\rm L_{X}\simeq 2\pm1\times 10^{38}}$ \ergps).
In addition, one source had no photons in the 0.5--7.0~keV 
band, suggesting that it was spurious and we therefore omitted it from our 
list. This left 73 sources on the S3 chip, of which 37 lay 
within the \dtwentyfive\ isophote of \src\ as given in \citet{devaucouleurs91},
which has semi-major and semi-minor axes of 2.3\arcmin\ and 0.7\arcmin,
respectively. We adjusted the position angle given in \citet{devaucouleurs91}
to 116\deg\ to agree with BC and match better the optical data. 
The source positions are given in Table~\ref{table_srclist}.

It is interesting to compare our detections with those made with \rosat.
Using the PSPC, 5 point sources were resolved  within the 
central ring (BC). 
There are bright \chandra\ detections (which we labelled
sources 1, 2 and 6 in Table~\ref{table_srclist}) within 15\arcsec\ 
($\sim$ the PSPC point-spread) of three of these sources. One of the \rosat\
sources would not fit on the S3 chip and the remaining source
was close to only one \chandra\ detection, source~71, which was too faint
(\lx $\sim 10^{38}$ \ergps) to have been seen with \rosat. This
indicates that the source seen with \rosat\ (which may, or may not, 
correspond with this \chandra\ detection) exhibited dramatic 
variability over the $\sim$ 6-year period between the observations.

For each source, lightcurves, spectra
and spectral responses were generated using standard \ciao\ tasks. 
Local background 
regions were chosen for each source to ensure that contamination from the 
diffuse
galaxy emission (or from the mild flaring) did not bias our results. These 
were accumulated from annuli centred on the source which were truncated so as 
to be contained entirely within the same \acis\ node. They were chosen to 
cover an area at least 8 times the extraction region, and containing at least
50 photons. We excluded from our background computation photons within
a region around each source obtained by scaling the axes of the appropriate
1-$\sigma$ encircled-energy ellipse by a factor 6 (we excluded such a large 
region since a few percent of a source's photons will leak from a region 
even as large as the 3-$\sigma$ encircled-energy ellipse, which is problematic
in the vicinity of bright sources).
To compensate for the progressive quantum-efficiency
degradation at low energies due to contamination on the optical blocking 
filter, we used the 
{\tt apply\_acisabs}\footnote{see \href{http://cxc.harvard.edu/ciao/threads/apply_acisabs}{http://cxc.harvard.edu/ciao/threads/apply\_acisabs}} script
to correct the ancillary response files. 
In order to investigate the possible extent of these
sources, dedicated software was used to generate radial brightness profiles
centred upon each source and excluding photons from the vicinity of the 
other detected point-sources. The profiles were
extracted initially in very fine, equally-spaced  bins 
($\sim$0.5\arcsec\ width), out to radii as large as 1\arcmin\ in order that
a reliable background could be estimated. For those sources close to the 
centre of the galaxy we adopted smaller radii (down to $\sim$0.5\arcmin) 
to minimize contamination from the diffuse gas,
or subsequently included a model to account for the diffuse component 
when fitting the profile (Sect~\ref{sect_srcextent}).

\section{Point source properties}

\subsection{Hardness ratios}

\begin{figure}
\plotone{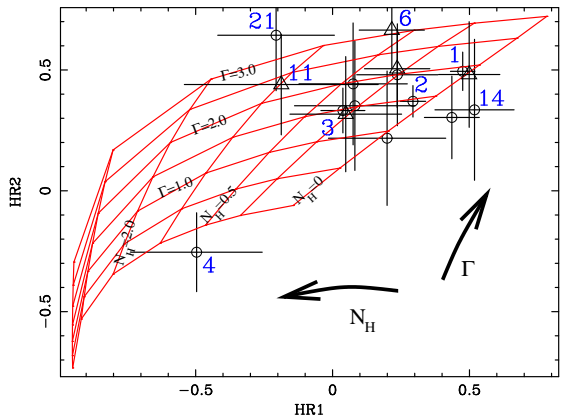}
\caption{Hardness ratios for all sources on the S3 chip with more than
40 counts. Overlaid is a grid of the loci of a simple absorbed power law 
model as $\Gamma$ and \nh\ vary. $\Gamma$ increases upwards,
with grid lines shown for $\Gamma$=0.0--3.0 in steps of 0.5. Several
representative lines are labelled.
\nh\ increases to the left and grid-lines are shown for 
\nh$=$0.0, 0.25, 0.5, 1.0, 2.0, 4.0 and 8.0 ${\rm \times 10^{22} cm^{-2}}$,
where \nh\ is column in excess of the appropriate Galactic value. 
Several lines of constant \nh\ are labelled on the figure in units of 
${\rm 10^{22} cm^{-2}}$. The identification numbers of a selection of interesting sources are indicated on the figure. Sources inside the \dtwentyfive\ region are shown as triangles, and those outside it are shown as circles.}
\label{fig_coco}
\end{figure}

In order to compare the spectral phenomenology of individual low-count sources 
it is convenient to construct hardness ratios 
\citep[\eg][]{white84,kim92,zezas02b}. 
Background-subtracted counts were measured 
for each source in three energy-bands, a soft-band (S) from 0.3--1.5~keV,
a medium-band (M) from 1.5--3.0~keV and a hard-band (H) from 3.0--5.0~keV,
and the hardness ratios HR1 and HR2 were computed. These are defined, 
respectively, as ${\rm HR1=(S-M)/(S+M)}$ and ${\rm HR2=(M-H)/(M+H)}$.
With these
definitions, HR1 is principally sensitive to the
hydrogen column-density in the line-of-sight, and HR2 depends upon the 
steepness of the spectrum, allowing us to construct a useful 
 ``colour-colour'' diagram (\ie\ HR2 {\em versus} HR1) for NGC\th 1332
(Fig~\ref{fig_coco}). 
In order to show how
the hardness ratios depend upon the underlying spectrum, we have 
overlaid a grid showing the loci of simple absorbed power law spectra
with a variety of different value of \nh\ and $\Gamma$. The position of
a given source relative to the grid has implications for its nature; 
most neutron-star or low-state black hole binaries, for example, 
would be expected to lie between the $\Gamma=1$ and the $\Gamma=2$ lines,
whereas high state black holes may be expected to lie above the 
$\Gamma=2.5$ line \citep{humphrey03a}. 

In NGC\th 1332, most of the sources are found to occupy regions 
consistent with \nh$\simeq$ Galactic \citep{dickey90} 
and $\Gamma \simeq$1--2, as seen in
other early-type galaxies. There is considerable scatter, but this merely
reflects the few photons detected in each source. Several sources 
show some evidence of intrinsic, or local, absorption in excess of the 
Galactic column but there are only two clear outliers (Sources 4 and 21).
These are confirmed by spectral-fitting; \eg\ Source~4 has 
\nh${\rm \simeq10^{22}\ cm^{-2}}$, in agreement with its hardness-plane
position. 
Sources 6 and 21 have rather soft spectra, although both sources are somewhat harder than the ``Supersoft'' and ``Quasisoft'' sources seen in other galaxies \citep{distefano03a}. Based on their colour-colour plane loci, they
may be consistent with high-state black hole systems, although this is only one possibility, which will need independent confirmation.
Individual spectral-fitting of these sources
does not allow the spectra to be constrained very well, but if free, the 
$\Gamma$ of a power law fit to each source spectrum does, indeed, rise to 
\gtsim 3.0, consistent with this picture. 

\subsection{Composite spectra}

\begin{figure}

\plotone{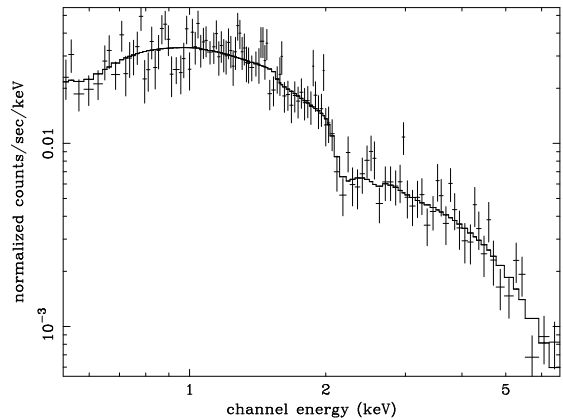}
\caption{Composite spectrum for all sources in the \dtwentyfive\ region, 
folded through the instrumental response,
and shown with the best-fit bremsstrahlung model.}
\end{figure}

In order to examine the overall properties of the source population,
and specifically to allow us to separate the unresolved source component
from the hot gas during spectral analysis of the diffuse 
emission (Paper II), we accumulated a 
composite background-subtracted spectrum of all 
the point-sources. Response matrices were generated for the data using the 
{\tt mkwarf} and {\tt mkrmf} \ciao\ tasks, and corrected for the 
effects of quantum efficiency degradation at low energies
analogously to the individual source spectra.
Since all of the brightest sources detected in \src\
were outside the \dtwentyfive\ region, we separately constructed a 
composite spectrum for only those sources within this region (thereby 
preventing our results from being unduly biased by the brightest objects).
We rebinned the data to ensure that in each bin there were at least 20
photons (to allow the use of the $\chi^2$ statistic) 
and a signal-to-noise ratio exceeding 3. 
Both spectra could be fitted excellently with simple one-component
bremsstrahlung or power law models, with \nh\ fixed at the nominal 
Galactic value, as shown in Table~\ref{table_compspectrum}.
Only for the power law fit to all the S3 sources
did freeing the \nh\ led to a modest improvement in the fit statistic. 
The spectra of the sources within the \dtwentyfive\ region and those on the entire chip were only slightly different.
For our fit of sources within the \dtwentyfive\ region,
our results were in excellent agreement with those of \citet{irwin03a},
consistent with the apparently ``universal'' nature of the composite spectra of
the lower-luminosity sources in early-type galaxies.

\subsection{Fluxing}
To measure the fluxes of the sources, we adopted two complementary techniques 
(which are effectively equivalent for the lower-counts sources). 
We rebinned the spectrum of each source to 
ensure at least 20 photons in each spectral bin 
(thereby allowing the use of $\chi^2$). 
When this gave fewer than 4 spectral bins, we instead regrouped
the spectrum into $\sim$12 bins of approximately equal width in energy
and adopted the Cash (maximum likelihood) fit statistic (determining the 
goodness-of-fit  from Monte-Carlo simulations).
Initially we fitted a simple power law model, the slope of which matched
the co-added source spectra within the \dtwentyfive\ region. 
When the ``null hypothesis probability'' (\ie\ the probability of the 
model fitting the data) was only acceptable at the 5\% level or less,
we rejected the model and allowed, separately, the 
photon index and the \nh\ to be free. In all cases we were then able to 
obtain statistically acceptable fits. Although more complex spectral 
models are generally more appropriate for Galactic LMXB
\citep{church01}, a simple power law can be used to provide an accurate
flux estimate in a relatively narrow energy-band and, for so few photons,
more complicated  models could not be  constrained.
To determine the flux errors, we modified the existing \xspec-models 
to redefine normalization as the flux within the 0.3--7.0~keV band. 
With this definition, flux errors could be obtained trivially without
correlations between different fit parameters biasing our results.
For most sources the data were consistent with the
co-added spectrum, although several of the brighter objects required 
additional \nh\ and in two cases we found an appreciably
softer spectrum. Whilst this method preferentially corrects the brightest
sources, only in a few cases was there a dramatic change in the \lx.
Considering the large luminosity errors of the faintest sources, this is 
unlikely to bias our results substantially.

In order to compare more directly with other authors, we additionally computed
counts-to-flux conversion factors for each of our sources. To allow for 
effective-area variations between detections, we folded the best-fit
composite source {\em bremsstrahlung} model 
(\nh=0.02$\times 10^{22}{\rm cm^{-2}}$, kT=8~keV) through the instrumental 
responses of each source to determine a local counts-to-flux 
conversion factor. The Poissonian errors upon the measured counts were
estimated using the Gehrels' approximation and translated into flux errors.
We found excellent
agreement between both methods of flux estimation, except in a few sources
which showed unusually absorbed spectra. The results of our flux estimation
are shown in Table~\ref{table_srclist}, which also defines our source naming
convention (sources being numbered in descending order of luminosity).

For the distance to \src\ (23 Mpc), we found that 8 sources
have \lx $>10^{39}$ \ergps, which qualifies them as ULX. The expected number
of bright background objects in this sample is, however, appreciable
(4, based on the Chandra deep field observations; Sect.~\ref{sect_xlf}), 
so some are undoubtedly background objects. 
In order to compare with other authors, we considered
separately only those ULX within the \dtwentyfive\ region, of which there
are 2, with an expected background number of 0.3. 
Neither of these sources has \lx$>2\times 10^{39}$ \ergps.

\subsection{Variability}

\begin{figure}
\plotone{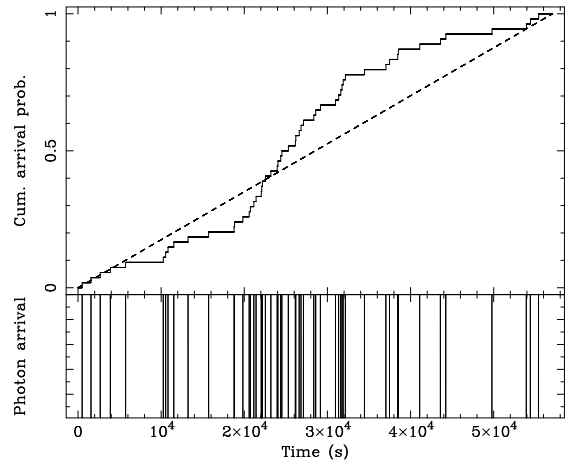}
\caption{Full energy-band ``lightcurve'' of the variable source~51. 
Since so few photons
were detected, we adopt a non-standard way of showing the lightcurve. In
the lower panel, each arriving photon is flagged as a vertical line,
whereas the upper panel shows a comparison between the measured 
cumulative arrival probability of the photons in the source (solid line)
and that expected for a non-varying source (dashed line)} \label{fig_variable}
\end{figure}

To search for variability we accumulated source and background lightcurves
with the maximum temporal resolution (3.24~s) for each source. 
Since the time binning was very fine 
(${\rm T_{bin}\ll T_{exp}}$), we were able to compare the observed 
distribution of  photon arrival times with that expected for a non-varying
source, using the  Kolmogorov-Smirnov (K-S) test (which requires 
{\em unbinned} data). The K-S test is fairly sensitive to large-amplitude 
variability occurring over timescales comparable to ${\rm T_{exp}}$.
To minimize false positive detections, since we were considering a total of 
74 sources (including the point-source in NGC\th 1331), we required that 
the null hypothesis (no variability) be rejected at 99.93\% significance 
(corresponding to a 5\% probability of a single false positive). 

We found that, of our sources, only two were variable at the required 
significance (as indicated in Table~\ref{table_srclist}). Since the
background is known not to be entirely constant (due to flaring), 
we tested these lightcurves having excluded flaring and found a comparably strong detection of variability. In addition, we
further tested the entire locally-accumulated background lightcurves  of these 
sources using the same test. In neither case
could the null hypothesis (no variability in the background) be 
rejected at better than 7\% significance. It is 
therefore highly improbable that the measured variability of the sources
arises from background fluctuations. 
Considering only the sources within
the \dtwentyfive\ region (and consequently lowering our detection significance
threshold) did not lead to any further detections of 
variability. Both variable sources had luminosities
$\sim$ 1--3~$\times 10^{38}$ \ergps, comparable to the Galactic ``Z-track'' 
sources, which can show variability over similar time-scales 
\citep{hasinger89}. There were too few photons from each source,
however, to classify these objects definitively. This will require a 
more sensitive, long-term study to track the spectral and timing behaviour.

\subsection{Spatial extent and the search for a central source}  \label{sect_srcextent}

\begin{figure*}
\plottwo{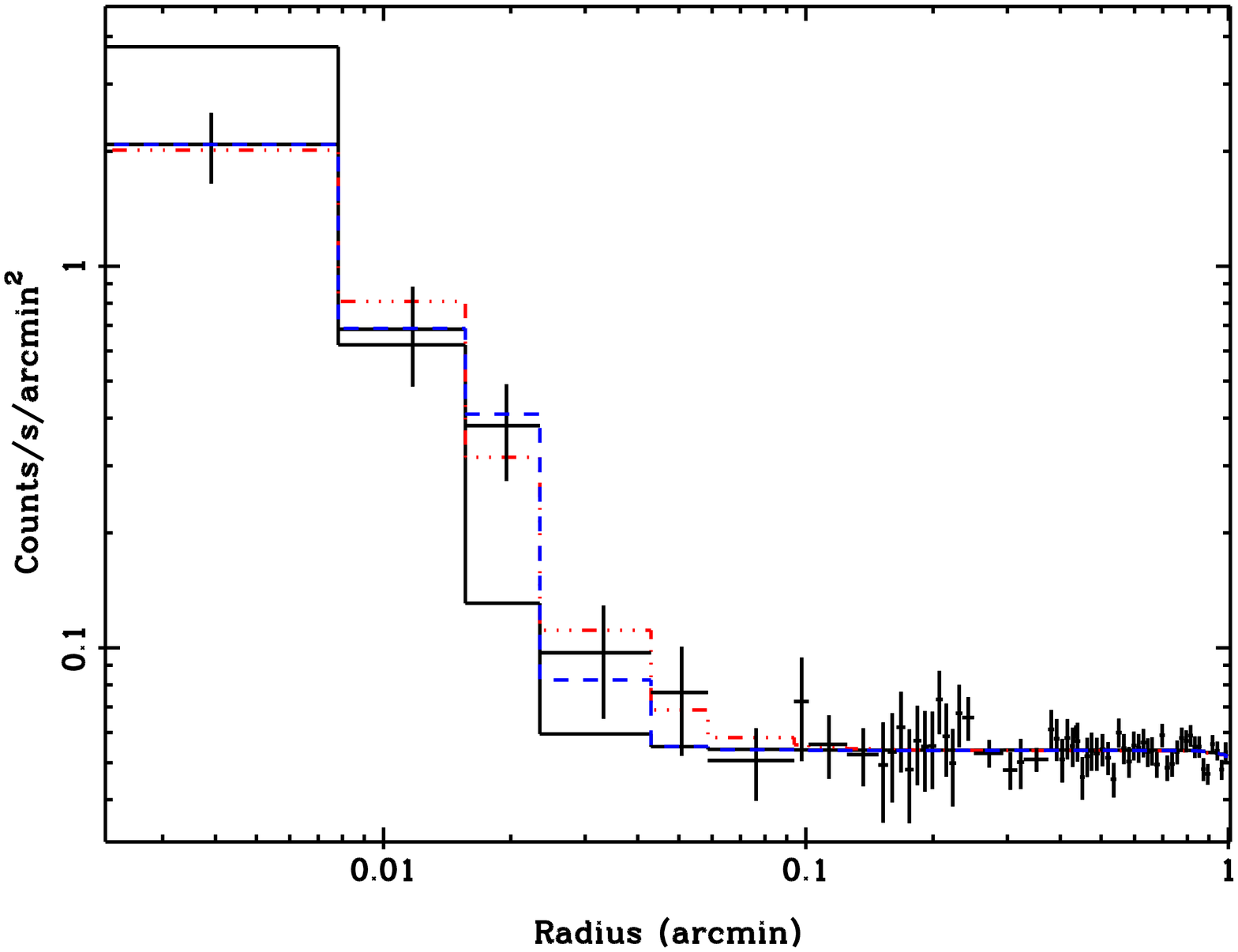}{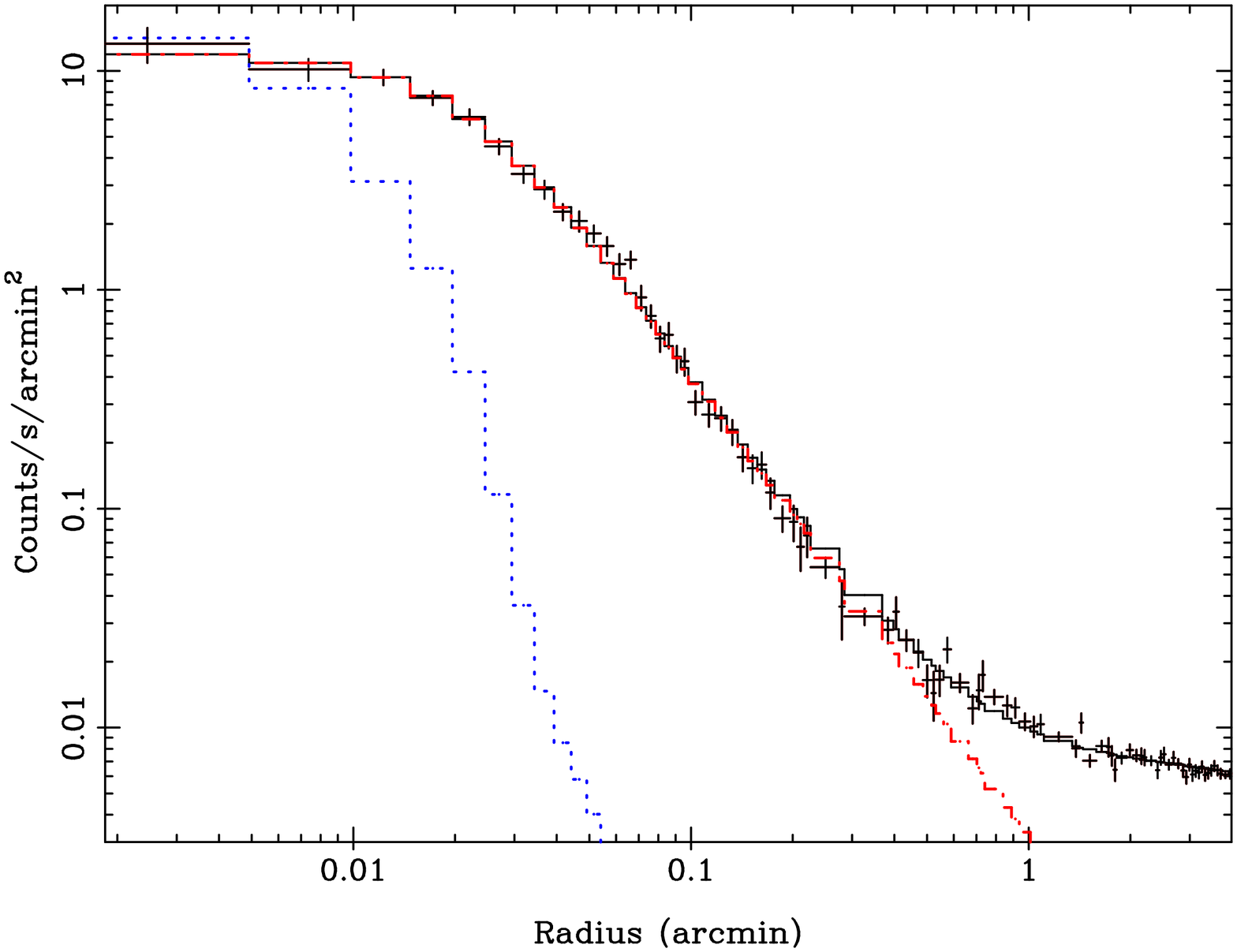}
\caption{Left panel: Full energy-band radial brightness profile for source~14. 
Three fits to the radial
brightness profile are also shown. These comprise a constant background term
and, respectively, the expected satellite point-spread 
computed at the peak source energy of 1.3~keV (solid line), a beta-model
(dash-dotted line) and a pair of point-sources separated by 1\arcsec\ 
(dashed line). Right panel: radial brightness profile of the diffuse central
emission from 0.3--7.0~keV. The data are shown with a fit comprising 
a constant background term plus a simple $\beta$-model. The $\beta$-model 
component is also shown separately (dash-dot line), and, for comparison, the azimuthally-averaged PSF (dotted line). There is no evidence of 
a central point-source.} \label{fig_radprofile}
\end{figure*}

The detection algorithm is capable of identifying both 
point-like and extended sources.
In order to identify any sources which may be extended, we examined the 
radial brightness profiles accumulated for each source. Fitting was 
performed using dedicated software which was able to make use of the radial
profiles described in Sect.~\ref{sect_detection} and to take account of 
the instrumental point-spread at the peak photon energy of each source.
The source
data were rebinned into $\sim$ 90 logarithmically-spaced radial bins
and then adjacent bins were regrouped to ensure none contained fewer 
than 20 photons.
The data were initially fitted with a model comprising a constant
background term and a point-source at the extraction centroid. 
To account for exposure variation over the S3 chip, an exposure-map was
generated for the source image at 1.7~keV and used to correct the applied
model. For those sources within $\sim$0.5\arcmin\ of the galaxy centroid,
the substantial diffuse emission contaminated the profile sufficiently that
a constant background model was no longer tenable. In these cases, we 
included an additional component, corresponding to a $\beta$-model (with 
parameters matching the diffuse galactic emission;
Paper III) and centred at an appropriate distance from the point-source.

In all except one case we found no evidence of spatial
extent greater than the instrumental PSF, placing limits of 
$\sim$50--100~pc (depending on the position in the field-of-view)
on the actual size of these objects. Fig.~\ref{fig_radprofile} shows the 
results for Source~14, which is clearly more extended than the 
PSF. Replacing the point-source with a $\beta$-model improved the 
$\chi^2$/dof from 104/60 to 59/58, which is significant at $>99.999$\%
(upon the basis of an f-test). We were also able to fit the radial profile
as two point-sources, one centred at the extraction centroid and one
${\rm 1.0^{+0.4}_{-0.06}}$\arcsec\ away from it, of approximately
equal brightness. Since we might expect two sources this far apart to
be distinguishable by \chandra, we tested this conclusion by 
simulating  an image of two-point sources (separated by 1\arcsec) at
an equivalent position in the focal-plane (generating 1.0~keV point-source 
images from the \caldb\ PSF hypercubes). 
If there were more than $\sim$ 100 counts in the sources they could
be resolved. When we  reduced the number of photons to $\sim$ 25 each 
(corresponding to the observed number of counts in Source~14), Poisson noise
blurred the outlines so that they remained unresolved.
This is a practical example of source-confusion.
The probability of a chance alignment of a background object 
within 1\arcsec\ of one of our detected sources was, however, only $\sim$1\%.
Such a low probability does tend to suggest that the source was genuinely
extended (having a size $\sim$ 1\arcsec), 
although we cannot altogether rule out the possibility of a chance
superposition. Future observations may be able to resolve the 
issue by identifying variability in this source, 
which would place limits on its size.

In addition, we searched  for a point-source embedded in the diffuse emission
at the galactic centre. The radial profile of the diffuse
emission out to 4.2\arcmin\ was grouped into 
$\sim$150 logarithmically-spaced bins.
These were regrouped to ensure at least 20 photons in
each bin. The data could be well-approximated with a single $\beta$-model
plus a constant background term (folding in the instrumental PSF at 1.0~keV
and correcting the model for exposure variations and gaps arising from 
removed point-sources), as shown in Fig.~\ref{fig_radprofile}. Although the
profile is extremely peaked (${\rm R_{core} = 1.2\pm0.1}$\arcsec)
its centre is significantly broader than the PSF of \acis\ 
at its position. Adding  a central point-source model does not improve 
the fit quality, placing an upper limit of 
13.4${\rm \times 10^{-4}\ count\ s^{-1}}$ 
(${\rm L_X<8.4\times 10^{38}}$ \ergps) on any central source. A more detailed
discussion of the morphology and radial profile of the diffuse emission
will be given in Paper III.

\subsection{Source identification}
In order to identify possible counterparts to the point-sources
we checked for objects listed in \ned, Simbad and the 
Tycho-2 catalogue of Galactic (foreground) stars 
\citep{hog00} within 2\arcsec\ of each source. We did not 
identify any obvious counterparts to X-ray sources which would enable us
independently to check the absolute astrometry. However, the centroid of the 
diffuse galactic emission of \src\ was located within 1\arcsec\ of the 
2MASS galaxy centroid (listed in \ned), which is accurate to $\sim$2\arcsec\ 
and there are no known pointing 
problems reported for our \chandra\ data. 
Therefore, we assumed that the astrometry of the detected sources was 
accurate to 2\arcsec. 
Only one counterpart was found to any source, that being \srctwo, the centroid
of which lies within 1\arcsec\ of an X-ray source (which we have not
listed in Table~\ref{table_srclist}). 

In addition, we checked for coincidence between the point source positions
and a list of GC candidates kindly supplied by A.\ Kundu (priv.\ comm.).
The GC candidates were identified in HST WFPC2 observations  centred on \src\ \citep{kundu01}, 
which encloses only $\sim$50\% of the \dtwentyfive\ region.
In order to perform this comparison it was
necessary to check the relative astrometry between \chandra\ 
and HST. A comparison of the X-ray centroid and the optical centre
in the appropriate archival
WFPC2 image clearly revealed a systematic offset of $\sim$1.6\arcsec\
between \chandra\ and HST. Compensating for this effect, we found that
9 sources lay within 0.75\arcsec\ of a GC candidate and so we assumed
they were associated with a globular cluster. We adopted this limit
since it corresponded to an apparent break in the distribution
of offsets between GC and the X-ray source positions. A total of 
30 X-ray sources coincided with the WFPC field-of-view 
(of which 23 were within \dtwentyfive), and there 
were a total of 204 GC candidates in this region of \src, so that 
$\sim$30\% of the X-ray sources within this region were identified
with GCs and 4.4\% of all GCs hosted LMXB. Both of these results
are consistent with observations of other early-type galaxies
\citep[\eg]{kundu02a,sarazin03}. We investigated whether there 
was any evidence of a systematic difference in the luminosities
of the GC and the non-GC sources by comparing the fluxes of the
two populations using a two-sample Kolmogorov-Smirnov test. 
We found no evidence of a statistically significant discrepancy 
between the two distributions, which agreed with
$\sim$60\% probability. The combined spectra of the GC sources exhibited no evidence of a systematic difference from the other sources.

\subsection{Spatial distribution of sources} \label{sect_spat_dist}

\begin{figure*}
\plottwo{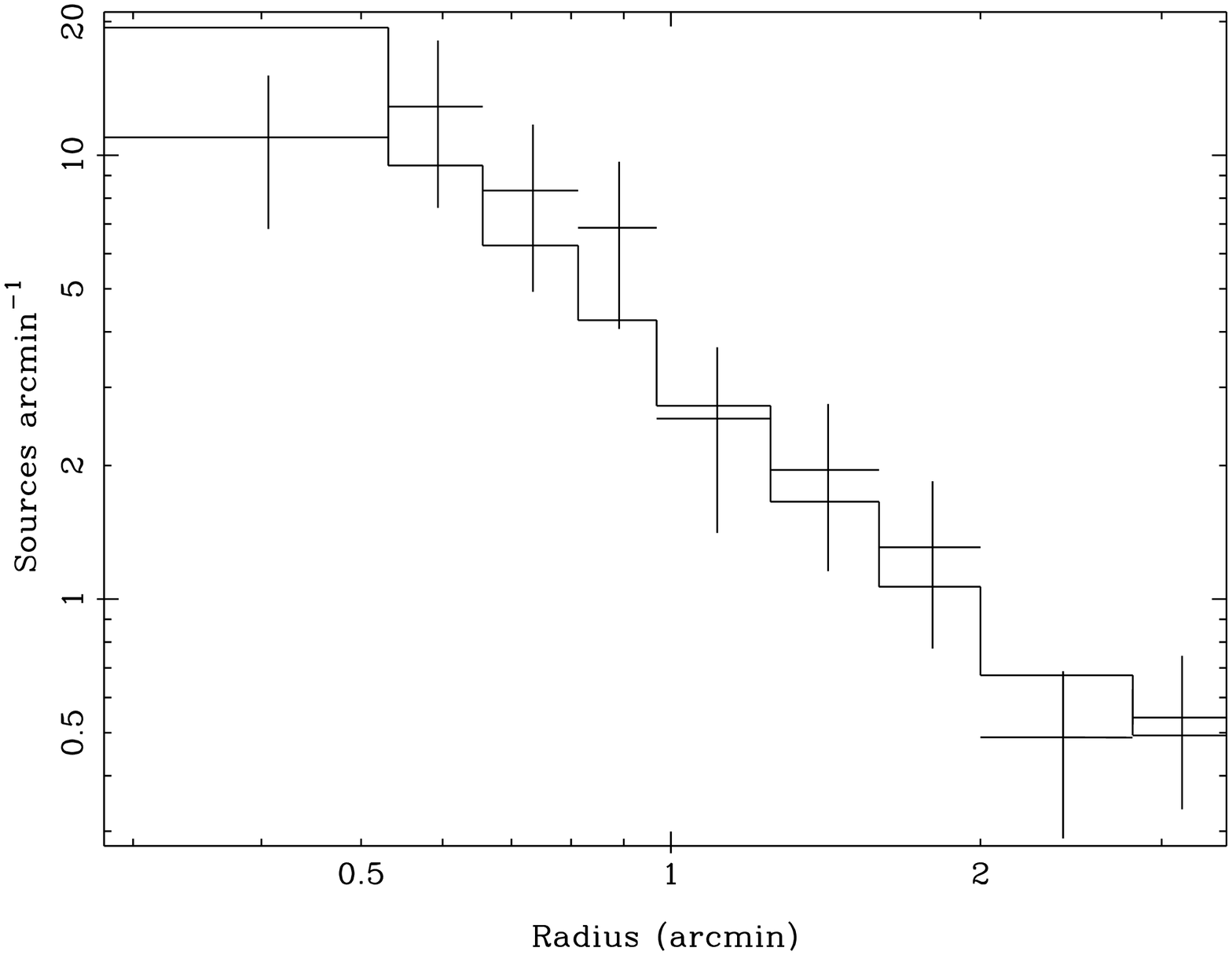}{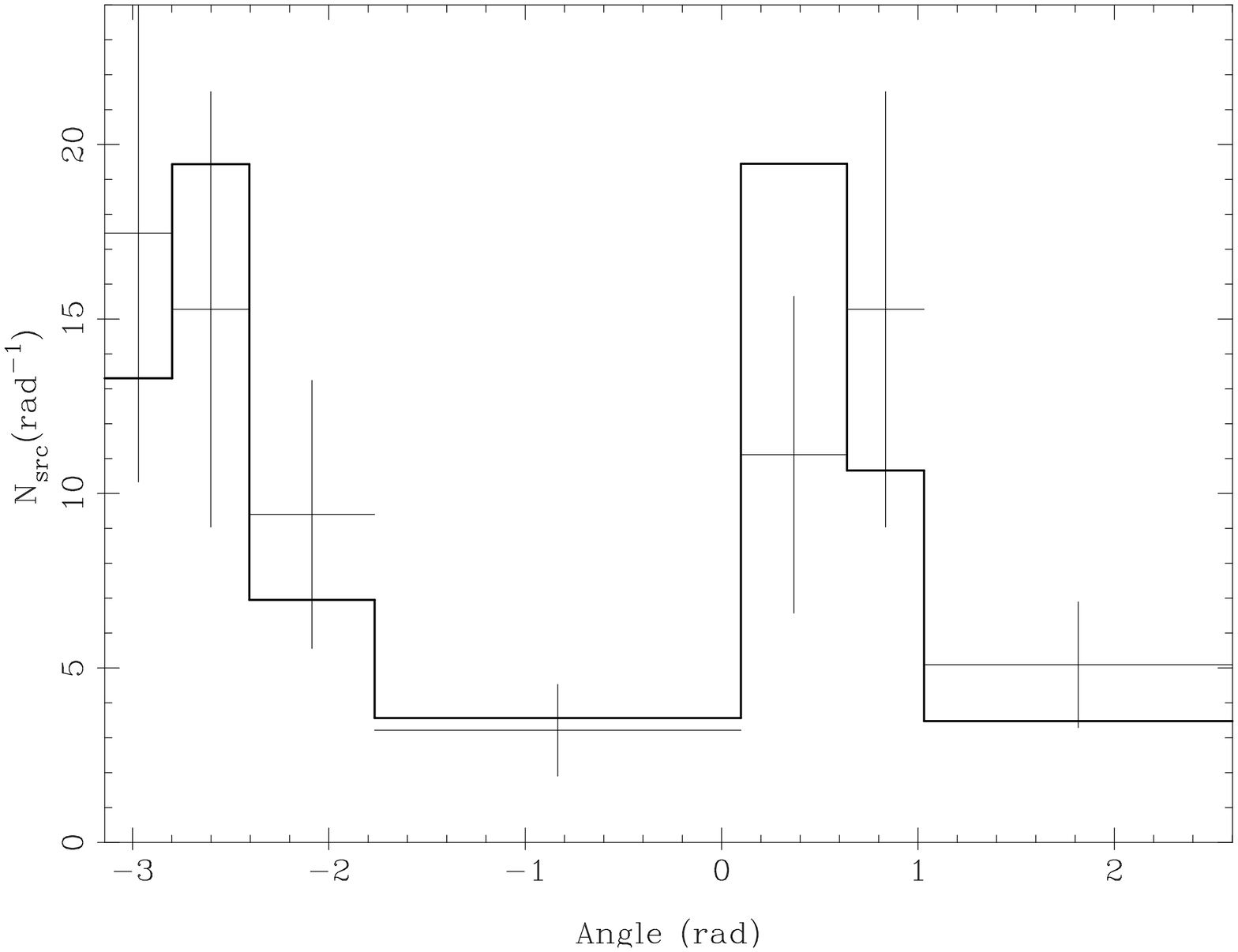}
\caption{Left panel: The radial distribution of the point-sources, binned 
to ensure at least 5 sources per bin and shown with the best-fit 
model following the optical light.
Right panel: The azimuthal distribution of sources, shown with the best-fit
model, which also follows the optical light.
} \label{fig_radial_dist}
\end{figure*}

We examined the spatial distribution of the point-sources in order to determine
to what extent they may follow either the optical light or the X-ray 
surface-brightness of the gas. In order to do this, we initially accumulated 
a histogram of the number of sources within each of 128 evenly-spaced 
radial bins 
centred at the galaxy centroid, and excluding the inner $\sim 0.2$\arcmin\
(where source detection incompleteness is most severe). 
In order to improve the signal-to-noise level,
the data were rebinned until each bin contained at least 5 sources.
Models were tested against the data using the Cash-statistic (the 
goodness-of-fit being determined by Monte-Carlo simulation). 
We also tried using
the $\chi^2$ statistic (which was less rigorous due to the low count-statistics)
and obtained comparable results. We tested two different models against the
data, one following the optical light of the galaxy and one following the 
X-ray contours. To approximate the optical profile it was convenient to fit a 
single de Vaucouleurs profile with its effective radius fixed to 
31\arcsec. We chose this value by correcting the semi-major axis effective 
radius measured by 
BC to account for the fact that our data were azimuthally averaged and not 
corrected for ellipticity. In addition, we tested 
a simple $\beta$-model, with a shape matching that of the diffuse emission,
against the data. 
In both cases,  we added an additional (constant) term 
to account for unrelated ``background'' sources.
We were able to obtain good fits to the data with both
models (null hypothesis probabilities of 56\% and 28\%, respectively), 
for which the inferred numbers of background sources agreed within errors.
Fig~\ref{fig_radial_dist} shows the data and the best-fit de Vaucouleurs model.
For our de Vaucouleurs fit, we found a total of $1300^{+1000}_{-800}$ 
background sources per degree, equivalent to a total of $1.9^{+1.5}_{-1.2}$
background sources in the \dtwentyfive\ region.
Taking our lowest measured flux to be the completeness limit, we would expect,
(based upon the number of sources detected by 
\citealt{tozzi01b} in the Chandra Deep Field image), there to be 
between $\sim$1400 and $\sim$2600 sources per degree (for the 
``soft'' and ``hard'' bands of these authors, corrected to our 0.3--7.0~keV 
band). Our data were clearly, therefore, in general agreement with the 
expected number of background sources from these observations.

In addition, we accumulated a histogram of the azimuthal source distribution.
We binned the data into 128 azimuthal bins, regrouping adjacent bins until
there were at least 5 sources in each.
We considered only  sources within 2\arcmin\ of the galaxy centroid 
to minimize  background-source contamination. Models 
were tested against the data using the Cash statistic, exactly as for
the radially-binned data. The data clearly showed azimuthal variation, as 
would be expected if the sources follow the optical light. Formally, we found
that a model in which the source density is constant with angle gives 
a null hypothesis only of 1.8\% and therefore was rejected. Alternatively,
for a model in which the number of sources followed the optical light,
the null hypothesis probability of 70\% indicated excellent agreement 
with the data (see Fig~\ref{fig_radial_dist}).
\subsection{Source completeness estimate}

\begin{figure}
\plotone{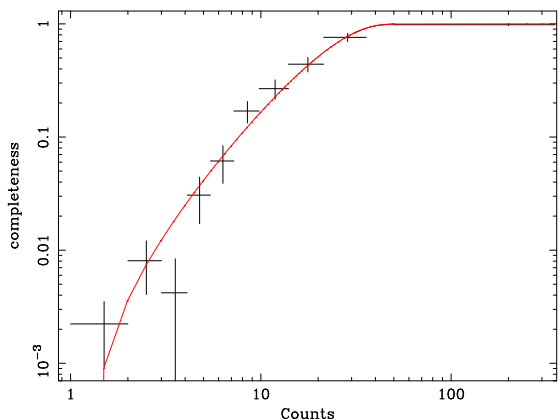}
\caption{Representative plot of the detection probability for a point-source at a given luminosity as a function of measured counts. Also shown is our polynomial approximation to the function.} \label{fig_completeness}
\end{figure}

A number of factors can influence the likelihood of detecting a given point
source, such as source confusion in crowded regions, increased local
background due to diffuse emission  and the degradation of the \chandra\ 
point-spread off-axis. 
The importance of correcting the measured XLF to account for incompleteness 
was demonstrated by \citet{kim03a}, 
who showed that failure to account for 
it can lend erroneous significance to XLF breaks. 
In order to estimate the level of incompleteness in our source detections
we adopted  a technique
similar to that outlined by \citet{kim03a}. We ran a large number of 
Monte-Carlo simulations in which a small number of sources were added to the 
{\em observed} image of \src\ in the 0.1--10~keV band and then measured the
fraction of these sources which could be found with the detection
algorithm.
Since our adoption of additional energy-bands did not appreciably add to the
number of detected sources, it was appropriate to consider only the 0.1--10~keV
energy-band in this calculation. The added sources were all assumed to have
spectra which were strongly peaked at 1~keV and appropriate images 
were extracted from the \caldb\ PSF hypercube and added to the image
(having been degraded with suitable Poisson noise).
The sources were added with an assumed luminosity function, 
${\rm (dN/dS)_0 \propto S^{-2}}$ (although its adopted shape 
does not affect the results) and, in order to speed the computation,
we added $\sim$ 10 sources at a time to each image down to a limiting flux
of 1 count per image. These were assumed to 
be distributed as the optical light (\ie\ an elliptical 
de-Vaucouleurs profile with axis ratio and alignment matching that of the 
\dtwentyfive\ ellipse and  with major-axis 0.55\arcsec; BC). 
Although ideally sources would be added one-at-a time 
so as not to introduce any biases, this expedient is not 
strictly necessary provided the incidence of overlap between added 
sources is low.
After a total of 400 simulations, we were able to measure the function
${\rm P_d(S)}$, which is the probability that a source with {\em exactly} 
S counts is detected. Due to the finite number of test sources, we binned 
them into a number of $\sim$logarithmically-spaced luminosity-bins and 
computed the fraction of sources detected within each bin
(Fig~\ref{fig_completeness}). 
${\rm P_d(S)}$ is given by the ratio of the measured and simulated 
differential XLF. We 
were then able to approximate ${\rm P_d(S)}$ by a fourth-order polynomial
in the range S=1--50 counts 
(also shown in Fig~\ref{fig_completeness}). We considered only 
simulated sources lying within the \dtwentyfive\ ellipse, since we 
later computed the XLF only in this region.  

\section{The X-ray luminosity function} \label{sect_xlf}

\begin{figure*}
\plotone{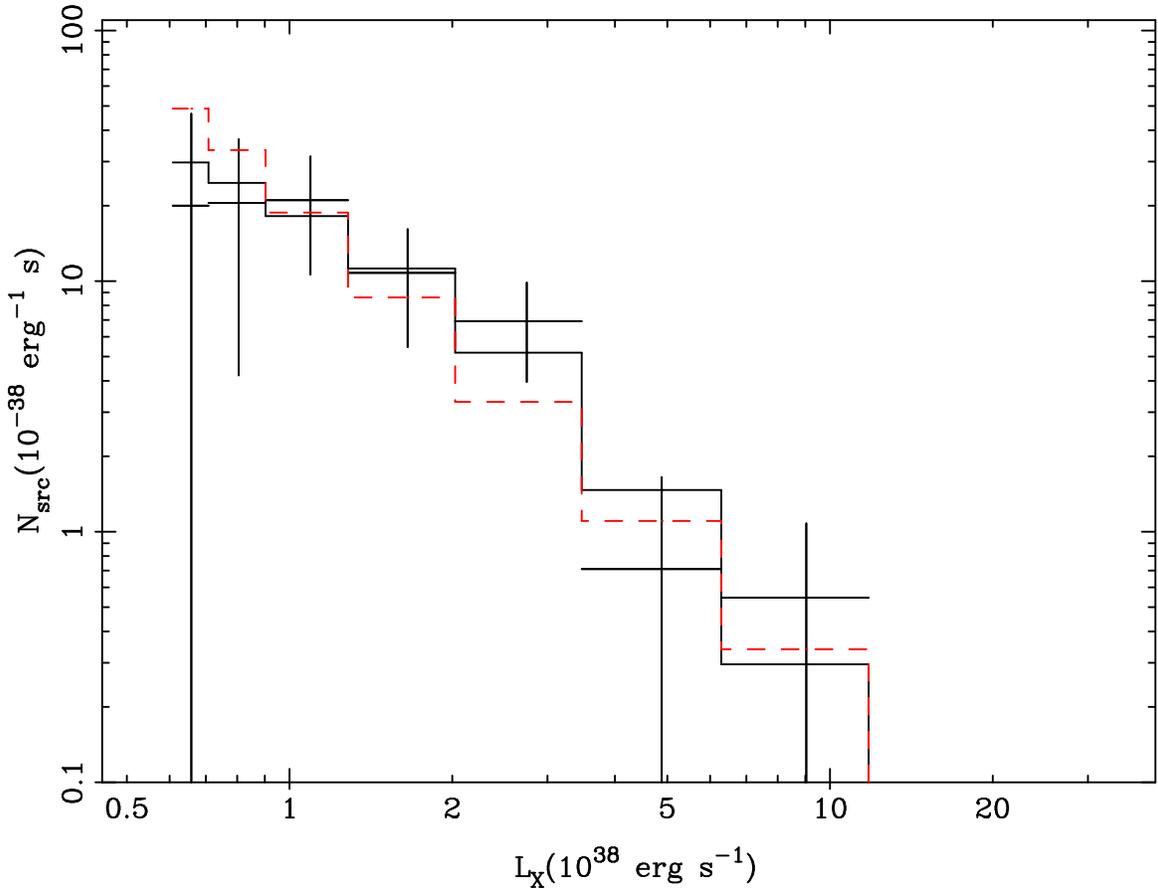}
\caption{The {\em differential} luminosity-function of the data. Error-bars,
estimated from the Gehrels' approximation, have been included on each point
only as a guide to the eye since the Cash statistic was used in fiting.
Also shown
are the best-fit completeness-corrected power law model (solid line)
and with the best-fit uncorrected simple power law model (dashed line). 
This demonstrates the impact of incompleteness on measuring the XLF.} 
\label{fig_lfunc}
\end{figure*}

In order to 
minimize the possible impact of background source contamination
(a strong source of potential error given field-to-field variations in 
the number of background objects), when computing the XLF 
we considered only those
sources within the \dtwentyfive\ region.
We fitted the differential luminosity function, separately considering 
the XLF derived  from our two different flux-estimation techniques.
Accordingly, the sources were binned as a function of luminosity in
such a way that the bin size increased geometrically (progressively 
scaling by a factor 2.0). We experimented with different scaling factors 
but found our results were relatively insensitive to the 
adopted binning.
Since our measurement included no upper luminosity cut-off,
it is significant that we detected no sources with luminosities 
greater than ${\rm \sim 12\times 10^{38}}$~\ergps. Therefore, an additional data-bin
containing zero sources and covering an extremely wide luminosity range
(from \lx${\rm \simeq 12}$--10000 ${\rm \times 10^{38}}$ \ergps) was 
added to our data (the fit statistic is extremely insensitive to the 
precise upper bound of this bin, provided it is larger than
$\sim 100\times 10^{38}$~\ergps).
Including this bin leads to steeper XLF fits than 
omitting it. To fit the XLF we minimized the Cash
statistic since our data were dominated by Poisson errors. 
The fit quality of the model was subsequently determine 
{\em via} Monte-Carlo simulations. 
To take account of flux measurement errors, we repeated the fit 100 times for 
each model, having first added appropriate Gaussian noise to the 
measured \lx\ of each source. If any resulting \lx\ would be 
below our lowest measured luminosity 
(assumed to be the source detection limit), it was recomputed. 
In order to test the reliability of this fitting procedure, we simulated 100 
luminosity functions from our best-fit model and attempted to recover the 
best-fit parameters. We found good agreement between the inferred and 
expected results and confirmed the expected statistical scatter, 
giving us confidence in our ability to determine parameters 
and estimate error-bars reliably, provided the simulations were representative, which we believe.

We tested a number of models against the data, in each case adding a 
separate component to account for the point-source population due to 
background sources. This term had a power law shape, \ie
\begin{equation}
{\rm \frac{dN}{dL} = K \left( \frac{L_X}{10^{38} erg\ s^{-1}} \right)^{-\beta}
}\end{equation}
We fixed  the slope, $\beta$, and the amplitude, K, to values consistent with 
the hard-band sources of \citet{tozzi01b}, \ie\ 
1.92 and 2.1 sources per $10^{38}$ \ergps, respectively
(correcting fluxes from the 2.0--10.0~keV band used by these authors to
our 0.3--7.0~keV band). This term did not contribute very significantly to
our fit and corresponds to $\sim$ 3.5
sources in total in the \dtwentyfive\
region. We found that allowing the normalization to be free, 
or adopting the shape and normalization appropriate for the 
soft band of \citet{tozzi01b} made little difference to our overall
fit. 
Without applying any form of completeness correction to our data,
we found that a single power law fit (with fluxes
estimated from spectral-fitting)
gave a very poor fit to the data (having a null hypothesis
probability of 0.3\%). Adding a break in the model  at
${\rm L_{break}=2.3^{+1.4}_{-0.7}\times 10^{38}}$ \ergps substantially 
improved the fit quality (giving a null hypothesis probability of 
42\%), 
although the power law slopes were poorly-constrained
(\eg\ ${\rm \beta=2.6^{+1.3}_{-0.7}}$ for \lx${\rm > L_{break}}$). 
To correct for source detection incompleteness 
we modified the simple power law by multiplication with our 
completeness estimate function (\ie\ we corrected the {\em model} rather
than the data, to allow us to continue using Poisson statistics). 
We obtained an excellent fit with the corrected single 
power law model (the null hypothesis probability now being 40\%).
The best-fit parameters were ${\rm \beta=2.3\pm0.4}$ and amplitude, 
K=$60\pm25$, and no break was required.
This indicates that incompleteness was responsible for the break measured
in the uncorrected data.
No further improvement in the fit could be obtained by adding a break to the 
power law model, and the resulting fit-parameters were very poorly-constrained.
The measured value of $\beta$ was in good agreement with  that obtained by \citet{kim03a} for a sample of 14 early-type galaxies. However, these authors computed source fluxes based on a counts-to-flux conversion, so this is not strictly a fair comparison.

We separately considered the XLF where fluxes were obtained from the 
count-rate, rather than the spectra. This
allowed us to compare our results more directly with other authors, who
tend to adopt this procedure. Exactly as above, we found that a 
broken power law gave a much better fit than a simple power law to the data,
but this effect vanished when proper account is made for the source
detection incompleteness. In this case, however, we obtained a somewhat
steeper slope for our luminosity-function (${\rm \beta=2.7\pm0.5}$;
K was correspondingly altered to ${\rm 87\pm33}$ ), 
although the two $\beta$ values agree within errors.
The differential XLF and the best-fit model are shown in Fig.~\ref{fig_lfunc},
along with the uncorrected single power law fit.
It is not surprising that
the XLF derived from fitting the spectra was less steep,
since the principal effect of the fitting was to uncover sources with
more than Galactic \nh, thereby populating more the higher 
luminosities.
Since some of these ``absorbed'' sources may have been background AGN
 and since the 
background XLF we adopted were computed from 
{\em counts-to-flux} conversion, it is probably more correct to adopt our
$\beta=2.7\pm0.5$ result. Intriguingly, this result 
is marginally ($\sim2.5\sigma$) steeper than the 
$\beta=2.0$ value found by 
\citet{kim03a} but is in good
agreement with their higher-luminosity slope when they fitted a 
broken power law to the combined, incompleteness-corrected XLF.
To verify that our slope was not biased by our use of data during the mild flaring, we recomputed all fluxes having excised data during the flares. The slope of the XLF was unchanged.

As a final note, we also attempted to fit the XLF using two alternative
fitting methods: a maximum-likelihood method appropriate for 
unbinned data \citep{crawford70} and minimization of  the  
Kolmogorov-Smirnov (K-S) test statistic. In both cases we obtained
less stringent constraints than for our binned-data analysis above. 
Since, by its nature, the K-S test is relatively 
insensitive at the sparsely-populated highest and lowest luminosities, 
that statistic in particular was much less reliable at determining the 
quality of the fit.

\section{Discussion}
We have found a total of 73 point-sources within the \chandra\ S3 field
of \src, including 37 which lie within the \dtwentyfive\ isophote and
none of which is coincident with the galaxy centroid. We expect
a total of $1.9^{+1.5}_{-1.2}$ unrelated background or foreground sources
to be within \dtwentyfive, so that the vast majority of these sources are
associated with the galaxy. 
The spatial distribution of the sources was consistent with the optical
isophotes, as may be expected from a population of sources of stellar 
origin which has not been recently disturbed by merger activity
\citep{zezas03a}. In total 30\% of the X-ray sources were associated 
with globular clusters, and 4.4\% of GCs contained an X-ray source,
both of which results are consistent with other early-type galaxies.
The phenomenology of the individual sources was 
consistent with expectations for an LMXB population. Two sources had 
soft spectra, consistent with high-state black-hole binaries and two
sources were variable during the observation. Intriguingly one source
was extended, although its radial profile was consistent with two
unresolved point-sources.

Without taking into account the impact of 
source detection incompleteness at the lowest luminosities (where 
source confusion and increased background due to diffuse gas reduce 
the probability of detecting a source),
we found that the XLF of the point-sources within 
\dtwentyfive\ can be fitted by a power law with a break at 
\lx$\sim 2\times10^{38}$ \ergps. Similar results have been reported in other 
early-type galaxies \citep[\eg][]{sarazin01,blanton01b}. However, when
we made corrections for the expected level of incompleteness, we 
found that the data were extremely well-fitted by a single power law,
with a differential slope, $\beta=$ $2.7\pm0.5$. 
Assuming that the XLF remains unbroken down to $10^{37}$ \ergps\
(as seen for sources in M\th 31: \citealt{kong02}), we
estimate a total luminosity of ${\rm 180\pm70 \times 10^{38}}$ \ergps\ due
to {\em unresolved} point-sources in the \dtwentyfive\ region.
Comparing this to the B-band luminosity of the galaxy, obtained from the face-on, extinction-corrected B-band luminosity listed in the \leda\footnote{\href{http://leda.univ-lyon1.fr/}{http://leda.univ-lyon1.fr/}} catalogue, we obtained  ${\rm L_{LMXB}/L_B=0.8\pm0.3\times 10^{30}}$~${\rm erg\ s^{-1} L_{B\odot}^{-1} }$. In addition, we compared
with the K-band luminosity (adopting the ${\rm K_{20}}$ magnitude obtained
by 2MASS), and obtained ${\rm L_{LMXB}/L_K=0.13\pm0.05\times 10^{30}}$~${\rm erg\ s^{-1} L_{K\odot}^{-1} }$.
Both of these values were in excellent agreement with observations of other
early-type galaxies \citep{kim03a}.

The slope of our XLF fit closely resembles that 
found at high luminosities in a variety of other early-type galaxies 
\citep{sarazin01,blanton01b,colbert03a}. A slope of 
$\beta \simeq 2.7$ is also strikingly similar to observations of a 
primarily old stellar population in M31, but which does not extend to 
such high luminosities \citep{kong03}. The slope is, however, 
marginally ($\sim2.5\sigma$)
steeper than the incompleteness-corrected {\em single} 
power law fits found in a recent survey of early-type galaxies \citep{kim03a}.
Intriguingly, these authors
reported some evidence of a break at $\sim 5\times 10^{38}$ \ergps, above
which $\beta \simeq 2.7$ is observed. A similar result has also been found 
for M84 \citep{finoguenov02}. A break at this luminosity should be 
observable within our data, and its apparent absence is 
intriguing. Whilst it is certainly true that {\em statistically}
acceptable single power law fits were found for most galaxies 
in the sample of \citet{kim03a}, the resulting slope
($\beta\simeq2.0$) may simply be an ``average'' of the slopes
above and below the break. The steep slope seen in \src\ might therefore 
suggest that such a break is absent in this galaxy. 
It must be stressed, however, that there are some differences in the analysis
procedures adopted by these authors and those used in the present work,
which may simply weight our fit somewhat more heavily towards the 
high luminosity sources (especially our inclusion of a high-luminosity, 
unoccupied bin).
Due to the small numbers of sources, however, the error-bars on the best-fit slope in \src\ are large, so that the disagreement is only at $\sim 2.5\sigma$. It remains to be seen whether comparable results are observed in any other galaxies containing significantly more X-ray point sources.
Nonetheless, if the absence of the break in our XLF is 
confirmed, it is an interesting result, which casts further doubt on
there being a ``universal'' break in early-type galaxies 
corresponding to the division between neutron-star and black-hole binary
systems.

The origin of the XLF break is still somewhat unclear. Although it was 
originally suggested that a break might mark the division between
neutron-star and black-hole binary systems \citep{sarazin01}, the
disparity between the reported break luminosities in several galaxies and
the Eddington limit for a 1.4${\rm M_\odot}$ (canonical) neutron-star 
suggests another origin, perhaps in the 
star-formation history of the galaxy \citep[\eg][]{jeltema03,wu01a}.
In that case, however, we might expect to see a strong correlation between
the age of the system (since the last major merger) and the XLF shape;
comparing the slopes found by \citet{kim03a}, including that for
NGC\th 720, and galaxy ages given by \citet{terlevich02}
we find no evidence of a correlation. However, this includes only 8 sources
and will need to be confirmed with a larger sample.

We detected  two ULX within the \dtwentyfive\ region of \src, although 
neither of these sources have \lx$>2\times 10^{39}$ \ergps. Such numbers 
are consistent with other  galaxies of similar morphology \citep{humphrey03a}.
The unbroken XLF of late-type galaxies suggests that the 
ULX associated with star-formation can be attributed to the high-luminosity
tail of the HMXB XLF, which is flatter than the LMXB XLF \citep{grimm03}.
It may be that, in a similar way, the ULX in early-type galaxies simply represent the high-\lx\ tail of the LMXB XLF.
\citet{irwin03b} have argued that sources with \lx$>2\times 10^{39}$ \ergps\
are extremely rare in  early-type galaxies, suggesting that the most
luminous ULX may therefore represent a population almost entirely alien to 
such galaxies. However luminous ULX have been reported in 
NGC\th 1399 and NGC\th 720  which are associated with GCs 
\citep[hence unlikely to be background objects;][]{angelini01,jeltema03}, 
so they are clearly
not entirely absent and therefore it is of interest to determine if their
paucity simply reflects the steepness of the XLF. In \src, since our fitting
took account of the absence of sources at high luminosities, the absence
of very luminous ULX is consistent with the high-luminosity tail of the XLF.
However, we also compared the ratio of the number of sources measured by 
\citet{irwin03b}
in the 1--2 ${\times 10^{39}}$ \ergps\ and the $>2\times 10^{39}$ band,
to that expected assuming our best-fit XLF slope ($\beta=2.7$).
We found a likelihood of $\sim 1$\% of reproducing the data
from the model, although allowing $\beta$ to steepen within its error-bars
increased the likelihood to $\sim 5$\%. Nonetheless this suggests 
the possibility of a 
change in the XLF at $\sim$1--2$\times 10^{39}$ \ergps. Intriguingly,
an XLF break was reported by \citet{jeltema03} around this luminosity.

\begin{acknowledgements}
The authors would like to thank Claude Canizares for helpful discussions
and carefully reading the manuscript. 
We would like to thank Arunav Kundu for very kindly supplying the 
list of GC candidates. We would also like to thank  Aaron Lewis for 
helpful discussions concerning \chandra\ analysis. 
This research has made use of the NASA/IPAC Extragalactic Database (\ned) 
which is operated by the Jet Propulsion Laboratory, California Institute of 
Technology, under contract with the National Aeronautics and Space 
Administration. This research also made use of the SIMBAD database,
operated at CDS, Strasbourg, France. 
In addition we used the LEDA on-line galaxy database. Support
for this work was provided by NASA through \chandra\ award number
G02-3104X, issued by the Chandra X-ray Observatory Center, which
is operated by the Smithsonian Astrophysical Observatory for and 
on behalf of NASA under contract NAS8-39073. Partial support for this
work was also provided by NASA under grant NAG5-13059, issued through 
the Office of Space Science Astrophysics Data Program.
\end{acknowledgements}

\bibliographystyle{apj_hyper}
\bibliography{paper_bibliography.bib}

\begin{thebibliography}{56}
\expandafter\ifx\csname natexlab\endcsname\relax\def\natexlab#1{#1}\fi

\bibitem[{{Angelini} {et~al.}(2001){Angelini}, {Loewenstein}, \&
  {Mushotzky}}]{angelini01}
\href{http://adsabs.harvard.edu/cgi-bin/nph-bib_query?bibcode=2001ApJ...557L..%
35A&db_key=AST}{{Angelini}, L., {Loewenstein}, M., \& {Mushotzky}, R.~F.} 2001,
  \apjl, 557, L35

\bibitem[{{Belczynski} {et~al.}(2003){Belczynski}, {Kalogera}, {Zezas}, \&
  {Fabbiano}}]{belczynski03}
\href{http://arxiv.org/abs/astro-ph/0310200}{{Belczynski}, K., {Kalogera}, V.,
  {Zezas}, A., \& {Fabbiano}, G.} 2003, \apjl, in press, astro-ph/0310200

\bibitem[{{Blanton} {et~al.}(2001){Blanton}, {Sarazin}, \&
  {Irwin}}]{blanton01b}
\href{http://adsabs.harvard.edu/cgi-bin/nph-bib_query?bibcode=2001ApJ...552..1%
06B&amp;db_key=AST}{{Blanton}, E.~L., {Sarazin}, C.~L., \& {Irwin}, J.~A.}
  2001, \apj, 552, 106

\bibitem[{{Brown} \& {Bregman}(2001)}]{brown01}
\href{http://adsabs.harvard.edu/cgi-bin/nph-bib_query?bibcode=2001ApJ...547..1%
54B&amp;db_key=AST}{{Brown}, B.~A. \& {Bregman}, J.~N.} 2001, \apj, 547, 154

\bibitem[{{Buote} \& {Canizares}(1996)}]{buote96a}
\href{http://adsabs.harvard.edu/cgi-bin/nph-bib_query?bibcode=1996ApJ...457..1%
77B&db_key=AST}{{Buote}, D.~A. \& {Canizares}, C.~R.} 1996, \apj, 457, 177

\bibitem[{{Buote} \& {Canizares}(1997)}]{buote97a}
\href{http://adsabs.harvard.edu/cgi-bin/nph-bib_query?bibcode=1997ApJ...474..6%
50B&db_key=AST}{{Buote}, D.~A. \& {Canizares}, C.~R.} 1997, \apj, 474, 650

\bibitem[{{Buote} {et~al.}(2004){Buote}, {Humphrey}, \& {Canizares}}]{buote04a}
{Buote}, D.~A., {Humphrey}, P.~J., \& {Canizares}, C.~R. 2004, in preparation.
  (Paper III)

\bibitem[{{Church} \& {Ba{\l}uci\'{n}ska-Church}(2001)}]{church01}
\href{http://adsabs.harvard.edu/cgi-bin/nph-bib_query?bibcode=2001A\%26A...369%
..915C&amp;db_key=AST}{{Church}, M.~J. \& {Ba{\l}uci\'{n}ska-Church}, M.} 2001,
  \aap, 369, 915

\bibitem[{{Colbert} {et~al.}(2003){Colbert}, {Heckman}, {Ptak}, \&
  {Strickland}}]{colbert03a}
\href{http://arxiv.org/abs/astro-ph/0305476}{{Colbert}, E., {Heckman}, T.,
  {Ptak}, A., \& {Strickland}, D.} 2003, \apj, in press, astro-ph/0305476

\bibitem[{{Colbert} \& {Mushotzky}(1999)}]{colbert99}
\href{http://adsabs.harvard.edu/cgi-bin/nph-bib_query?bibcode=1999ApJ...519...%
89C&amp;db_key=AST}{{Colbert}, E.~J.~M. \& {Mushotzky}, R.~F.} 1999, \apj, 519,
  89

\bibitem[{{Colbert} \& {Ptak}(2002)}]{colbert02}
\href{http://cdsads.u-strasbg.fr/cgi-bin/nph-bib_query?bibcode=2002ApJS..143..%
.25C&amp;db_key=AST}{{Colbert}, E.~J.~M. \& {Ptak}, A.~F.} 2002, \apjs, 143, 25

\bibitem[{{Crawford} {et~al.}(1970){Crawford}, {Jauncey}, \&
  {Murdoch}}]{crawford70}
\href{http://adsabs.harvard.edu/cgi-bin/nph-bib_query?bibcode=1970ApJ...162..4%
05C&amp;db_key=AST}{{Crawford}, D.~F., {Jauncey}, D.~L., \& {Murdoch}, H.~S.}
  1970, \apj, 162, 405

\bibitem[{{Davis} \& {Mushotzky}(2003)}]{davis03}
\href{http://arxiv.org/abs/astro-ph/0312211}{{Davis}, D.~S. \& {Mushotzky},
  R.~F.} 2003, \apj, in press, astro-ph/0312211

\bibitem[{{de Vaucouleurs} {et~al.}(1991){de Vaucouleurs}, {de Vaucouleurs},
  {Corwin}, {Buta}, {Paturel}, \& {Fouque}}]{devaucouleurs91}
\href{http://adsabs.harvard.edu/cgi-bin/nph-bib_query?bibcode=1991trcb.book...%
..D&amp;db_key=AST}{{de Vaucouleurs}, G., {de Vaucouleurs}, A., {Corwin},
  H.~G.}, {Buta}, R.~J., {Paturel}, G., \& {Fouque}, P. 1991, {Third Reference
  Catalogue of Bright Galaxies} (Volume 1-3, XII, 2069 pp.~7 figs..~
  Springer-Verlag Berlin Heidelberg New York)

\bibitem[{{Dickey} \& {Lockman}(1990)}]{dickey90}
\href{http://adsabs.harvard.edu/cgi-bin/nph-bib_query?bibcode=1990ARA\%26A..28%
..215D&amp;db_key=AST}{{Dickey}, J.~M. \& {Lockman}, F.~J.} 1990, \araa, 28,
  215

\bibitem[{{DiStefano} \& {Kong}(2003)}]{distefano03a}
\href{http://arxiv.org/abs/astro-ph/0311374}{{DiStefano}, R. \& {Kong}, A.}
  2003, \apj, submitted, astro-ph/0311374

\bibitem[{{Fabbiano}(1989)}]{fabbiano89}
\href{http://adsabs.harvard.edu/cgi-bin/nph-bib_query?bibcode=1989ARA\%26A..27%
...87F&db_key=AST}{{Fabbiano}, G.} 1989, \araa, 27, 87

\bibitem[{{Fabbiano} \& {White}(2003)}]{fabbiano03}
\href{http://arxiv.org/abs/astro-ph/0307077}{{Fabbiano}, G. \& {White}, N.~E.}
  2003, in Compact Stellar X-ray Sources, ed. W.~H.~G. {Lewin} \& M.~{van der
  Klis} (C.U.P.), astro-ph/0307077

\bibitem[{{Finoguenov} \& {Jones}(2002)}]{finoguenov02}
\href{http://adsabs.harvard.edu/cgi-bin/nph-bib_query?bibcode=2002ApJ...574..7%
54F&amp;db_key=AST}{{Finoguenov}, A. \& {Jones}, C.} 2002, \apj, 574, 754

\bibitem[{{Freeman} {et~al.}(2002){Freeman}, {Kashyap}, {Rosner}, \&
  {Lamb}}]{freeman02}
\href{http://cdsads.u-strasbg.fr/cgi-bin/nph-bib_query?bibcode=2002ApJS..138..%
185F&amp;db_key=AST}{{Freeman}, P.~E., {Kashyap}, V., {Rosner}, R., \& {Lamb},
  D.~Q.} 2002, \apjs, 138, 185

\bibitem[{{Grimm} {et~al.}(2003){Grimm}, {Gilfanov}, \& {Sunyaev}}]{grimm03}
\href{http://adsabs.harvard.edu/cgi-bin/nph-bib_query?bibcode=2003MNRAS.339..7%
93G&amp;db_key=AST}{{Grimm}, H.-J., {Gilfanov}, M., \& {Sunyaev}, R.} 2003,
  \mnras, 339, 793

\bibitem[{{Hasinger} \& {van der Klis}(1989)}]{hasinger89}
\href{http://adsabs.harvard.edu/cgi-bin/nph-bib_query?bibcode=1989A\%26A...225%
...79H&amp;db_key=AST}{{Hasinger}, G. \& {van der Klis}, M.} 1989, \aap, 225,
  79

\bibitem[{{H{\o}g} {et~al.}(2000){H{\o}g}, {Fabricius}, {Makarov}, {Urban},
  {Corbin}, {Wycoff}, {Bastian}, {Schwekendiek}, \& {Wicenec}}]{hog00}
\href{http://adsabs.harvard.edu/cgi-bin/nph-bib_query?bibcode=2000A\%26A...355%
L..27H&amp;db_key=AST}{{H{\o}g}, E., {Fabricius}, C., {Makarov}, V.~V.,
  {Urban}, S.}, {Corbin}, T., {Wycoff}, G., {Bastian}, U., {Schwekendiek}, P.,
  \& {Wicenec}, A. 2000, \aap, 355, L27

\bibitem[{{Humphrey} {et~al.}(2004){Humphrey}, {Buote}, \&
  {Canizares}}]{humphrey04b}
{Humphrey}, P.~J., {Buote}, D.~A., \& {Canizares}, C.~R. 2004, in preparation.
  (Paper II)

\bibitem[{{Humphrey} {et~al.}(2003){Humphrey}, {Fabbiano}, {Elvis}, {Church},
  \& {Ba{\l}uci{\' n}ska-Church}}]{humphrey03a}
\href{http://adsabs.harvard.edu/cgi-bin/nph-bib_query?bibcode=2003MNRAS.344..1%
34H&amp;db_key=AST}{{Humphrey}, P.~J., {Fabbiano}, G., {Elvis}, M., {Church},
  M.~J.}, \& {Ba{\l}uci{\' n}ska-Church}, M. 2003, \mnras, 344, 134

\bibitem[{{Irwin} {et~al.}(2003{\natexlab{a}}){Irwin}, {Athey}, \&
  {Bregman}}]{irwin03a}
\href{http://adsabs.harvard.edu/cgi-bin/nph-bib_query?bibcode=2003ApJ...587..3%
56I&db_key=AST}{{Irwin}, J.~A., {Athey}, A.~E., \& {Bregman}, J.~N.}
  2003{\natexlab{a}}, \apj, 587, 356

\bibitem[{{Irwin} {et~al.}(2003{\natexlab{b}}){Irwin}, {Bregman}, \&
  {Athey}}]{irwin03b}
\href{http://arxiv.org/abs/astro-ph/0312393}{{Irwin}, J.~A., {Bregman}, J.~N.,
  \& {Athey}, A.~E.} 2003{\natexlab{b}}, \apj, in press, astro-ph/0312393

\bibitem[{{Jeltema} {et~al.}(2003){Jeltema}, {Canizares}, {Buote}, \&
  {Garmire}}]{jeltema03}
\href{http://adsabs.harvard.edu/cgi-bin/nph-bib_query?bibcode=2003ApJ...585..7%
56J&db_key=AST}{{Jeltema}, T.~E., {Canizares}, C.~R., {Buote}, D.~A., \&
  {Garmire}, G.~P.} 2003, \apj, 585, 756

\bibitem[{{Kilgard} {et~al.}(2002){Kilgard}, {Kaaret}, {Krauss}, {Prestwich},
  {Raley}, \& {Zezas}}]{kilgard02}
\href{http://adsabs.harvard.edu/cgi-bin/nph-bib_query?bibcode=2002ApJ...573..1%
38K&amp;db_key=AST}{{Kilgard}, R.~E., {Kaaret}, P., {Krauss}, M.~I.,
  {Prestwich}, A.~H.}, {Raley}, M.~T., \& {Zezas}, A. 2002, \apj, 573, 138

\bibitem[{{Kim} \& {Fabbiano}(2003{\natexlab{a}})}]{kim03b}
\href{http://adsabs.harvard.edu/cgi-bin/nph-bib_query?bibcode=2003ApJ...586..8%
26K&amp;db_key=AST}{{Kim}, D. \& {Fabbiano}, G.} 2003{\natexlab{a}}, \apj, 586,
  826

\bibitem[{{Kim} \& {Fabbiano}(2003{\natexlab{b}})}]{kim03a}
\href{http://arxiv.org/abs/astro-ph/0312104}{{Kim}, D.-W. \& {Fabbiano}, G.}
  2003{\natexlab{b}}, \apj, submitted, astro-ph/0312104

\bibitem[{{Kim} {et~al.}(1992){Kim}, {Fabbiano}, \& {Trinchieri}}]{kim92}
\href{http://adsabs.harvard.edu/cgi-bin/nph-bib_query?bibcode=1992ApJS...80..6%
45K&amp;db_key=AST}{{Kim}, D.-W., {Fabbiano}, G., \& {Trinchieri}, G.} 1992,
  \apjs, 80, 645

\bibitem[{{King}(2002)}]{king02}
\href{http://adsabs.harvard.edu/cgi-bin/nph-bib_query?bibcode=2002MNRAS.335L..%
13K&amp;db_key=AST}{{King}, A.~R.} 2002, \mnras, 335, L13

\bibitem[{{King} {et~al.}(2001){King}, {Davies}, {Ward}, {Fabbiano}, \&
  {Elvis}}]{king01}
\href{http://adsabs.harvard.edu/cgi-bin/nph-bib_query?bibcode=2001ApJ...552L.1%
09K&db_key=AST}{{King}, A.~R., {Davies}, M.~B., {Ward}, M.~J., {Fabbiano}, G.},
  \& {Elvis}, M. 2001, \apjl, 552, L109

\bibitem[{{Kong} {et~al.}(2003){Kong}, {DiStefano}, {Garcia}, \&
  {Greiner}}]{kong03}
\href{http://adsabs.harvard.edu/cgi-bin/nph-bib_query?bibcode=2003ApJ...585..2%
98K&amp;db_key=AST}{{Kong}, A.~K.~H., {DiStefano}, R., {Garcia}, M.~R., \&
  {Greiner}, J.} 2003, \apj, 585, 298

\bibitem[{{Kong} {et~al.}(2002){Kong}, {Garcia}, {Primini}, {Murray}, {Di
  Stefano}, \& {McClintock}}]{kong02}
\href{http://adsabs.harvard.edu/cgi-bin/nph-bib_query?bibcode=2002ApJ...577..7%
38K&amp;db_key=AST}{{Kong}, A.~K.~H., {Garcia}, M.~R., {Primini}, F.~A.,
  {Murray}, S.~S.}, {Di Stefano}, R., \& {McClintock}, J.~E. 2002, \apj, 577,
  738

\bibitem[{{Kraft} {et~al.}(2001){Kraft}, {Kregenow}, {Forman}, {Jones}, \&
  {Murray}}]{kraft01}
\href{http://adsabs.harvard.edu/cgi-bin/nph-bib_query?bibcode=2001ApJ...560..6%
75K&amp;db_key=AST}{{Kraft}, R.~P., {Kregenow}, J.~M., {Forman}, W.~R.,
  {Jones}, C.}, \& {Murray}, S.~S. 2001, \apj, 560, 675

\bibitem[{{Kundu} {et~al.}(2002){Kundu}, {Maccarone}, \& {Zepf}}]{kundu02a}
\href{http://adsabs.harvard.edu/cgi-bin/nph-bib_query?bibcode=2002ApJ...574L..%
.5K&db_key=AST}{{Kundu}, A., {Maccarone}, T.~J., \& {Zepf}, S.~E.} 2002, \apjl,
  574, L5

\bibitem[{{Kundu} {et~al.}(2003){Kundu}, {Maccarone}, {Zepf}, \&
  {Puzia}}]{kundu03}
\href{http://cdsads.u-strasbg.fr/cgi-bin/nph-bib_query?bibcode=2003ApJ...589L.%
.81K&amp;db_key=AST}{{Kundu}, A., {Maccarone}, T.~J., {Zepf}, S.~E., \&
  {Puzia}, T.~H.} 2003, \apjl, 589, L81

\bibitem[{{Kundu} \& {Whitmore}(2001)}]{kundu01}
\href{http://adsabs.harvard.edu/cgi-bin/nph-bib_query?bibcode=2001AJ....122.12%
51K&amp;db_key=AST}{{Kundu}, A. \& {Whitmore}, B.~C.} 2001, \aj, 122, 1251

\bibitem[{{Makishima} {et~al.}(2000){Makishima}, {Kubota}, {Mizuno}, {Ohnishi},
  {Tashiro}, {Aruga}, {Asai}, {Dotani}, {Mitsuda}, {Ueda}, {Uno}, {Yamaoka},
  {Ebisawa}, {Kohmura}, \& {Okada}}]{makishima00}
\href{http://cdsads.u-strasbg.fr/cgi-bin/nph-bib_query?bibcode=2000ApJ...535..%
632M&amp;db_key=AST}{{Makishima}, K., {et~al.}} 2000, \apj, 535, 632

\bibitem[{{Matsushita} {et~al.}(1994){Matsushita}, {Makishima}, {Awaki},
  {Canizares}, {Fabian}, {Fukazawa}, {Loewenstein}, {Matsumoto}, {Mihara},
  {Mushotzky}, {Ohashi}, {Ricker}, {Serlemitsos}, {Tsuru}, {Tsusaka}, \&
  {Yamazaki}}]{matsushita94}
\href{http://adsabs.harvard.edu/cgi-bin/nph-bib_query?bibcode=1994ApJ...436L..%
41M&db_key=AST}{{Matsushita}, K., {et~al.}} 1994, \apjl, 436, L41

\bibitem[{{Paczy\'{n}ski}(1983)}]{paczynski83}
\href{http://cdsads.u-strasbg.fr/cgi-bin/nph-bib_query?bibcode=1983ApJ...267..%
315P&amp;db_key=AST}{{Paczy\'{n}ski}, B.} 1983, \apj, 267, 315

\bibitem[{{Roberts} \& {Warwick}(2000)}]{roberts00}
\href{http://cdsads.u-strasbg.fr/cgi-bin/nph-bib_query?bibcode=2000MNRAS.315..%
.98R&amp;db_key=AST}{{Roberts}, T.~P. \& {Warwick}, R.~S.} 2000, \mnras, 315,
  98

\bibitem[{{Sarazin} {et~al.}(2001){Sarazin}, {Irwin}, \& {Bregman}}]{sarazin01}
\href{http://adsabs.harvard.edu/cgi-bin/nph-bib_query?bibcode=2001ApJ...556..5%
33S&amp;db_key=AST}{{Sarazin}, C.~L., {Irwin}, J.~A., \& {Bregman}, J.~N.}
  2001, \apj, 556, 533

\bibitem[{{Sarazin} {et~al.}(2003){Sarazin}, {Kundu}, {Irwin}, {Sivakoff},
  {Blanton}, \& {Randall}}]{sarazin03}
\href{http://cdsads.u-strasbg.fr/cgi-bin/nph-bib_query?bibcode=2003ApJ...595..%
743S&amp;db_key=AST}{{Sarazin}, C.~L., {Kundu}, A., {Irwin}, J.~A., {Sivakoff},
  G.~R.}, {Blanton}, E.~L., \& {Randall}, S.~W. 2003, \apj, 595, 743

\bibitem[{{Saviane} {et~al.}(2003){Saviane}, {Hibbard}, \& {Rich}}]{saviane03}
\href{http://arxiv.org/abs/astro-ph/0311200}{{Saviane}, I., {Hibbard}, J.~E.,
  \& {Rich}, R.~M.} 2003, \apj, in press, astro-ph/0311200

\bibitem[{{Terlevich} \& {Forbes}(2002)}]{terlevich02}
\href{http://adsabs.harvard.edu/cgi-bin/nph-bib_query?bibcode=2002MNRAS.330..5%
47T&db_key=AST}{{Terlevich}, A.~I. \& {Forbes}, D.~A.} 2002, \mnras, 330, 547

\bibitem[{{Tonry} {et~al.}(2001){Tonry}, {Dressler}, {Blakeslee}, {Ajhar},
  {Fletcher}, {Luppino}, {Metzger}, \& {Moore}}]{tonry01}
\href{http://adsabs.harvard.edu/cgi-bin/nph-bib_query?bibcode=2001ApJ...546..6%
81T&db_key=AST}{{Tonry}, J.~L., {Dressler}, A., {Blakeslee}, J.~P., {Ajhar},
  E.~A.}, {Fletcher}, A.~., {Luppino}, G.~A., {Metzger}, M.~R., \& {Moore},
  C.~B. 2001, \apj, 546, 681

\bibitem[{{Tozzi} {et~al.}(2001){Tozzi}, {Rosati}, {Nonino}, {Bergeron},
  {Borgani}, {Gilli}, {Gilmozzi}, {Hasinger}, {Grogin}, {Kewley}, {Koekemoer},
  {Norman}, {Schreier}, {Szokoly}, {Wang}, {Zheng}, {Zirm}, \&
  {Giacconi}}]{tozzi01b}
\href{http://adsabs.harvard.edu/cgi-bin/nph-bib_query?bibcode=2001ApJ...562...%
42T&amp;db_key=AST}{{Tozzi}, P., {et~al.}} 2001, \apj, 562, 42

\bibitem[{{White} \& {Marshall}(1984)}]{white84}
\href{http://cdsads.u-strasbg.fr/cgi-bin/nph-bib_query?bibcode=1984ApJ...281..%
354W&amp;db_key=AST}{{White}, N.~E. \& {Marshall}, F.~E.} 1984, \apj, 281, 354

\bibitem[{{White} {et~al.}(1995){White}, {Nagase}, \& {Parmar}}]{white95}
{White}, N.~E., {Nagase}, F., \& {Parmar}, A.~N. 1995, in X-ray Binaries, ed.
  W.~H.~G. {Lewin}, J.~{van Paradijs}, \& E.~P.~J. {van den Heuvel} (C.U.P.),
  1--57

\bibitem[{{Wu}(2001)}]{wu01a}
\href{http://adsabs.harvard.edu/cgi-bin/nph-bib_query?bibcode=2001PASA...18..4%
43W&db_key=AST}{{Wu}, K.} 2001, Publications of the Astronomical Society of
  Australia, 18, 443

\bibitem[{{Zezas} \& {Fabbiano}(2002)}]{zezas02d}
\href{http://adsabs.harvard.edu/cgi-bin/nph-bib_query?bibcode=2002ApJ...577..7%
26Z&amp;db_key=AST}{{Zezas}, A. \& {Fabbiano}, G.} 2002, \apj, 577, 726

\bibitem[{{Zezas} {et~al.}(2002){Zezas}, {Fabbiano}, {Rots}, \&
  {Murray}}]{zezas02b}
\href{http://adsabs.harvard.edu/cgi-bin/nph-bib_query?bibcode=2002ApJS..142..2%
39Z&amp;db_key=AST}{{Zezas}, A., {Fabbiano}, G., {Rots}, A.~H., \& {Murray},
  S.~S.} 2002, \apjs, 142, 239

\bibitem[{{Zezas} {et~al.}(2003){Zezas}, {Hernquist}, {Fabbiano}, \&
  {Miller}}]{zezas03a}
\href{http://arxiv.org/abs/astro-ph/0310567}{{Zezas}, A., {Hernquist}, L.,
  {Fabbiano}, G., \& {Miller}, J.} 2003, \apjl, in press, astro-ph/0310567

\end{thebibliography}


\begin{deluxetable*}{llllllll}
\tablecaption{Sources detected \label{table_srclist}}
\tablehead {
\colhead{Src} & \colhead{Name} & \colhead{Rate} & \colhead{${\rm L_X^{spec}}$} & 
\colhead{${\rm L_X^{rate}}$} & \colhead{${\rm \Delta R}$} & \colhead{${\rm In\ D_{25}}$} &\colhead{Comment}  \\
\colhead{}  & \colhead{} &\colhead{(${\rm 10^{-4}\th s^{-1}}$)}&\colhead{(${\rm 10^{38}erg\ s^{-1}}$)}&\colhead{(${\rm 10^{38}erg\ s^{-1}}$)}&\colhead{($\prime$)}& \colhead{} & \colhead{}  \\ 
}
\startdata 
1 &{CXOU J032636.5-211809} &$81.$ &$58.\pm 5.$ &$60.\pm 5.$ &$4.9$ &no &  R \\
2 &CXOU J032614.8-212126 &$95.$ &$53.\pm 4.$ &$46.\pm 3.$ &$1.4$ &no & A, R, NG  \\
3 &CXOU J032604.4-212258 &$45.$ &$36.\pm 5.$ &$24.\pm 3.$ &$4.1$ &no & A \\
4 &CXOU J032620.7-212151 &$11.$ &$23.^{+9.}_{-3.}$ &$5.3\pm 1.3$ &$1.9$ &no & A, NG \\
5 &CXOU {J032632.8-212306}&$6.0$ &$12.^{+4.}_{-3.}$ &$8.6\pm 3.6$ &$4.7$ &no & {A, C}\\
6 &CXOU J032618.1-212014 &$18.$ &$12.^{+3.}_{-2.}$ &$12.\pm 2.$ &$0.23$ &yes & A, B, R, NG  \\
7 &CXOU J032601.7-211726 &$21.$ &$12.\pm 2.$ &$12.\pm 2.$ &$4.5$ &no &  \\
8 &CXOU J032616.7-211954 &$19.$ &$12.\pm 2.$ &$11.\pm 2.$ &$0.26$ &yes & NG \\
9 &CXOU J032618.7-212031 &$17.$ &$7.8\pm 1.8$ &$8.4\pm 1.6$ &$0.51$ &yes & NG \\
10 &CXOU J032606.8-212016 &$12.$ &$7.2^{+2.0}_{-1.3}$ &$6.1\pm 1.4$ &$2.4$ &no & \\
11 &CXOU {J032616.9-212006}&$8.8$ &$6.7\pm 2.6$ &$5.3\pm 2.3$ &$0.094$ &yes & NG \\
12 &CXOU J032612.5-212342 &$6.2$ &$6.6^{+2.1}_{-1.8}$ &$3.3\pm 1.2$ &$3.7$ &no & A \\
13 &CXOU J032609.7-212211 &$9.1$ &$5.9\pm 1.9$ &$4.6\pm 1.3$ &$2.7$ &no & \\
14 &CXOU {J032614.6-211903} &$10.$ &$5.8\pm 1.6$ &$5.2\pm 1.3$ &$1.2$ &no & E \\
15 &CXOU {J032615.2-212001} &$9.3$ &$5.6^{+2.7}_{-0.9}$ &$5.6\pm 1.5$ &$0.50$ &yes & \\
16 &CXOU J032622.6-212418 &$8.5$ &$5.5^{+2.0}_{-1.2}$ &$4.7\pm 1.4$ &$4.3$ &no & \\
17 &CXOU J032615.2-211924 &$2.9$ &$4.9^{+2.1}_{-1.6}$ &$1.5\pm 0.8$ &$0.87$ &yes & A \\
18 &CXOU J032617.1-212011 &$2.8$ &$4.5^{+2.1}_{-1.6}$ &$1.9\pm 2.2$ &$0.082$ &yes & G \\
19 &CXOU J032617.6-212013 &$2.8$ &$4.5^{+2.0}_{-1.6}$ &$2.0\pm 2.4$ &$0.13$ &yes & G \\
20 &CXOU J032622.8-211917 &$7.3$ &$4.4^{+1.8}_{-0.9}$ &$3.7\pm 1.1$ &$1.5$ &no & \\
21 &CXOU J032603.1-211921 &$7.4$ &$4.3^{+2.3}_{-0.7}$ &$3.7\pm 1.1$ &$3.4$ &no & A, B  \\
22 &CXOU J032631.3-212339 &$5.8$ &$3.9\pm 1.8$ &$3.2\pm 1.2$ &$4.8$ &no & \\
23 &CXOU J032620.0-212023 &$5.7$ &$3.4\pm 1.3$ &$3.0\pm 1.1$ &$0.69$ &yes & NG\\
24 &CXOU J032616.3-211958 &$4.7$ &$3.4\pm 1.8$ &$2.7\pm 1.4$ &$0.27$ &yes & G\\
25 &CXOU J032622.4-212132 &$7.1$ &$3.2^{+1.6}_{-0.8}$ &$3.5\pm 1.1$ &$1.8$ &no & NG\\
26 &CXOU J032613.1-212020 &$4.7$ &$3.0^{+1.3}_{-0.8}$ &$2.3\pm 0.9$ &$0.99$ &yes & NG\\
27 &CXOU J032618.7-211814 &$5.9$ &$3.0^{+1.5}_{-0.7}$ &$3.1\pm 1.1$ &$1.9$ &no & V \\
28 &CXOU J032605.7-212058 &$5.3$ &$2.7^{+1.1}_{-0.9}$ &$2.6\pm 1.00$ &$2.8$ &no & \\
29 &CXOU J032619.0-212030 &$4.3$ &$2.4\pm 0.9$ &$2.2\pm 0.9$ &$0.54$ &yes & G \\
30 &CXOU J032619.4-212021 &$3.8$ &$2.4\pm 1.0$ &$2.0\pm 1.0$ &$0.53$ &yes & G \\
31 &CXOU J032619.3-212052 &$3.6$ &$2.3^{+0.7}_{-1.1}$ &$1.8\pm 0.8$ &$0.88$ &yes &  NG\\
32 &CXOU J032619.9-212019 &$3.0$ &$2.3^{+1.1}_{-0.9}$ &$1.9\pm 1.00$ &$0.64$ &yes & NG\\
33 &CXOU J032616.4-211953 &$4.4$ &$2.3^{+1.2}_{-0.7}$ &$2.2\pm 1.0$ &$0.31$ &yes & G\\
34 &CXOU J032601.6-212021 &$3.6$ &$2.3^{+1.1}_{-0.8}$ &$1.8\pm 0.9$ &$3.6$ &no & \\
35 &CXOU J032625.5-211622 &$2.1$ &$2.3^{+1.5}_{-0.9}$ &$1.8\pm 1.1$ &$4.2$ &no & {C}\\
36 &CXOU J032618.3-212011 &$3.9$ &$2.3^{+1.5}_{-0.9}$ &$2.4\pm 1.2$ &$0.25$ &yes & NG\\
37 &CXOU J032621.6-211930 &$3.5$ &$2.2^{+0.8}_{-0.9}$ &$1.7\pm 0.8$ &$1.2$ &no & NG\\
38 &CXOU J032627.4-212235 &$5.7$ &$2.2^{+1.4}_{-0.8}$ &$2.9\pm 1.2$ &$3.4$ &no & \\
39 &CXOU J032616.2-211953 &$4.3$ &$2.1^{+1.2}_{-0.7}$ &$2.2\pm 1.0$ &$0.34$ &yes & \\
40 &CXOU J032615.5-211941 &$4.8$ &$2.1^{+1.1}_{-0.7}$ &$2.4\pm 1.0$ &$0.60$ &yes & \\
41 &CXOU J032602.2-212140 &$3.7$ &$2.0^{+1.3}_{-0.9}$ &$2.0\pm 1.2$ &$3.8$ &no & \\
42 &CXOU J032620.6-211713 &$3.6$ &$1.9^{+1.3}_{-0.5}$ &$1.8\pm 0.8$ &$3.0$ &no & \\
43 &CXOU J032613.9-211944 &$3.6$ &$1.8^{+1.4}_{-0.5}$ &$1.8\pm 0.8$ &$0.87$ &yes & \\
44 &CXOU J032610.3-211842 &$3.1$ &$1.8^{+0.9}_{-0.7}$ &$1.6\pm 0.8$ &$2.1$ &no & \\
45 &CXOU J032620.1-211958 &$2.3$ &$1.7^{+0.9}_{-0.7}$ &$1.2\pm 0.7$ &$0.67$ &yes & NG\\
46 &CXOU J032620.4-212112 &$3.0$ &$1.6^{+0.9}_{-0.7}$ &$1.5\pm 0.8$ &$1.3$ &yes & NG\\
47 &CXOU J032631.3-212210 &$3.2$ &$1.6^{+1.1}_{-0.7}$ &$1.6\pm 0.9$ &$3.8$ &no & \\
48 &CXOU J032624.2-212040 &$2.0$ &$1.6^{+0.7}_{-0.9}$ &$1.2\pm 0.8$ &$1.7$ &yes & \\
49 &CXOU J032620.6-212317 &$2.2$ &$1.5^{+1.0}_{-0.7}$ &$1.2\pm 0.8$ &$3.2$ &no & \\
50 &CXOU J032630.9-211824 &$3.3$ &$1.5^{+0.9}_{-0.6}$ &$1.7\pm 0.8$ &$3.6$ &no & \\
51 &CXOU J032606.7-212139 &$3.0$ &$1.4^{+1.2}_{-0.5}$ &$1.5\pm 0.8$ &$2.9$ &no & V \\
52 &CXOU J032621.9-212023 &$3.3$ &$1.3^{+1.3}_{-0.5}$ &$2.1\pm 1.00$ &$1.1$ &yes & NG\\
53 &CXOU J032614.6-211955 &$1.7$ &$1.3^{+1.0}_{-0.5}$ &$0.93\pm 0.70$ &$0.65$ &yes & \\
\enddata
\end{deluxetable*}

\setcounter{table}{0}
\begin{deluxetable*}{llllllll}
\tablecaption{Sources detected (contd.)}
\tablehead {
\colhead{Src} & \colhead{Name} & \colhead{Rate} & \colhead{${\rm L_X^{spec}}$} & 
\colhead{${\rm L_X^{rate}}$} & \colhead{${\rm \Delta R}$} & \colhead{${\rm In\ D_{25}}$} &\colhead{Comment}  \\
\colhead{}  & \colhead{} &\colhead{(${\rm 10^{-4}\th s^{-1}}$)}&\colhead{(${\rm 10^{38}erg\ s^{-1}}$)}&\colhead{(${\rm 10^{38}erg\ s^{-1}}$)}&\colhead{($\prime$)}& \colhead{} & \colhead{}  \\ 
}
\startdata 
54 &CXOU J032620.5-212041 &$2.5$ &$1.3^{+0.7}_{-0.6}$ &$1.3\pm 0.7$ &$0.94$ &yes & G \\
55 &CXOU {J032613.9-212005} &$2.1$ &$1.2^{+1.1}_{-0.4}$ &$1.1\pm 0.7$ &$0.79$ &yes &  \\
56 &CXOU J032625.1-212043 &$2.0$ &$1.1^{+1.0}_{-0.4}$ &$1.1\pm 0.7$ &$1.9$ &yes & \\
57 &CXOU J032621.8-211930 &$2.6$ &$1.1^{+0.5}_{-0.7}$ &$1.3\pm 0.7$ &$1.2$ &no & NG\\
58 &CXOU J032609.2-212156 &$1.5$ &$1.1^{+0.5}_{-0.7}$ &$0.81\pm 0.66$ &$2.6$ &no & \\
59 &CXOU J032622.8-211620 &$1.4$ &$1.1^{+0.8}_{-0.5}$ &$0.76\pm 0.64$ &$4.0$ &no & \\
60 &CXOU J032623.1-211945 &$1.6$ &$1.1^{+0.7}_{-0.5}$ &$0.79\pm 0.60$ &$1.4$ &no & NG\\
61 &CXOU J032616.5-211920 &$1.7$ &$1.0\pm 0.6$ &$0.83\pm 0.63$ &$0.81$ &yes & \\
62 &CXOU J032634.7-211755 &$2.0$ &$0.96^{+1.2}_{-0.36}$ &$1.1\pm 0.7$ &$4.6$ &no & \\
63 &CXOU J032615.4-211948 &$2.1$ &$0.96^{+1.1}_{-0.34}$ &$1.0\pm 0.7$ &$0.53$ &yes & \\
64 &CXOU J032612.8-211960 &$2.1$ &$0.93^{+1.1}_{-0.33}$ &$1.1\pm 0.7$ &$1.0$ &yes & \\
65 &CXOU J032615.7-212038 &$2.6$ &$0.93\pm 0.57$ &$1.3\pm 0.8$ &$0.63$ &yes & NG\\
66 &CXOU J032620.8-211957 &$1.4$ &$0.88^{+0.84}_{-0.40}$ &$0.73\pm 0.61$ &$0.83$ &yes & G \\
67 &CXOU J032621.5-212029 &$1.3$ &$0.82^{+0.63}_{-0.49}$ &$0.68\pm 0.62$ &$1.1$ &yes & G \\
68 &CXOU J032616.6-211927 &$1.6$ &$0.79^{+0.97}_{-0.31}$ &$0.79\pm 0.65$ &$0.69$ &yes & \\
69 &CXOU J032604.7-211613 &$1.7$ &$0.78^{+0.60}_{-0.46}$ &$0.89\pm 0.66$ &$4.8$ &no & \\
70 &CXOU J032607.9-211828 &$1.5$ &$0.75^{+0.63}_{-0.43}$ &$0.77\pm 0.61$ &$2.7$ &no & \\
71 &CXOU J032618.8-212148 &$1.4$ &$0.63^{+0.56}_{-0.38}$ &$0.69\pm 0.60$ &$1.7$ &no & R(?), NG\\
72 &CXOU J032614.2-211950 &$1.4$ &$0.63^{+0.57}_{-0.45}$ &$0.75\pm 0.67$ &$0.76$ &yes & \\
73 &CXOU J032615.7-212017 &$1.2$ &$0.34^{+0.55}_{-0.28}$ &$0.61\pm 0.59$ &$0.39$ &yes & NG\\
\enddata 
\tablecomments{Source list for all point-sources detected on the \acis\ S3 chip, 
excluding the source coincident with \srctwo. Two luminosity estimates are given,
${\rm L_X^{spec}}$ and ${\rm L_X^{rate}}$ which are, respectively, the luminosity
estimated from the spectra and from the count-rate (exposure-corrected).
${\rm \Delta}$R is the 
distance of the source from the galaxy centroid (in arcmin). 
{The count-rate column quotes count-rates {\em not} corrected for exposure-map variations.}
The comment
abbreviations are as follows: A--- absorbed spectrum, B--- {soft spectrum}, {C--- close to chip edge (flux may be a unreliable)}, E--- extended, G--- coincident with a globular cluster, 
NG--- in the WFPC2 fov but not coincident with a GC,
R--- also seen with \rosat\ and V---variable.
}
\end{deluxetable*} 

\begin{deluxetable}{lllll}
\tablecaption{Composite source spectra \label{table_compspectrum}}
\tablehead{ 
\colhead{Model} & \colhead{${\rm \chi^2/dof}$} &  
\colhead{\nh} & 
\colhead{${\rm \Gamma}$} & \colhead{${\rm kT}$} \\
\colhead{}  & \colhead{}  & \colhead{${\rm 10^{22}\ cm^{-2}}$}  &\colhead{}&
\colhead{(keV)} \\ 
}
\startdata 
\multicolumn{5}{l}{All sources} \\ \hline
power law & 122/119 & 0.0223 & $1.40\pm 0.05$& \\
          & 115.3/118 & $0.07\pm0.03$ & $1.5\pm0.1$ &  \\
bremsstrahlung & 115/119 & 0.0223 & & $15^{+6}_{-4}$\\
               & 114/118 & $0.03\pm0.02$ & & $12^{+7}_{-3}$ \\ \hline
\multicolumn{5}{l}{Sources in \dtwentyfive\ only} \\ \hline
power law & 50.0/46 & 0.0223 & $1.55\pm 0.10$& \\
          & 50.0/45 & $<0.075$ & $1.6\pm0.1$ &  \\
bremsstrahlung & 50.8/46 & 0.0223 & & $8^{+4}_{-2}$\\
               & 49.2/45 & $<0.031$ & & $10^{+5}_{-3}$ \\ 
\enddata 
\tablecomments{Results of fitting the composite spectra of all the point
sources on the \acis-S3 chip and, separately, only those sources in the
\dtwentyfive\ region. \nh\ values shown include the Galactic column
density, which was 0.0223${\rm \times 10^{22} cm^{-2}}$ 
\citep{dickey90}. Where no error-bars are shown
the parameter was fixed. Errors are  90\% confidence intervals.}
\end{deluxetable}

\end{document}